\documentclass[aps,prd,amssymb,amsmath,eqsecnum,nofootinbib,twocolumn]{revtex4-2}
\usepackage{xcolor,ulem}
\usepackage{scalerel}
\usepackage{algorithm}
\usepackage{mathrsfs}
\usepackage[a4paper,total={7in,9in}]{geometry}

\usepackage{color}
\usepackage{bm}
\usepackage{ulem}
\usepackage{graphicx}
\usepackage{comment}
\graphicspath{{./img/}}
\usepackage{scalerel}
\usepackage{tikz}
\usepackage{upgreek}
\usetikzlibrary{svg.path}

\definecolor{orcidlogocol}{HTML}{A6CE39}
\tikzset{
  orcidlogo/.pic={
    \fill[orcidlogocol] svg{M256,128c0,70.7-57.3,128-128,128C57.3,256,0,198.7,0,128C0,57.3,57.3,0,128,0C198.7,0,256,57.3,256,128z};
    \fill[white] svg{M86.3,186.2H70.9V79.1h15.4v48.4V186.2z}
                 svg{M108.9,79.1h41.6c39.6,0,57,28.3,57,53.6c0,27.5-21.5,53.6-56.8,53.6h-41.8V79.1z M124.3,172.4h24.5c34.9,0,42.9-26.5,42.9-39.7c0-21.5-13.7-39.7-43.7-39.7h-23.7V172.4z}
                 svg{M88.7,56.8c0,5.5-4.5,10.1-10.1,10.1c-5.6,0-10.1-4.6-10.1-10.1c0-5.6,4.5-10.1,10.1-10.1C84.2,46.7,88.7,51.3,88.7,56.8z};
  }
}

\newcommand\orcidicon[1]{\href{https://orcid.org/#1}{\mbox{\scalerel*{
\begin{tikzpicture}[yscale=-1,transform shape]
\pic{orcidlogo};
\end{tikzpicture}
}{|}}}}

\usepackage{hyperref}

\begin{document}

\title{Conformal compactification and affine-null  metric formulation of the Einstein equations}

\author{Thomas M\"adler \orcidicon{0000-0001-5076-3362}}
\email{thomas.maedler._.at._.mail.udp.cl}
\affiliation{Escuela de Obras Civiles and Instituto de Estudios Astrof\'isicos, Facultad de Ingenier\'ia y Ciencias, Universidad Diego Portales, Av. Ej\'ercito Libertador 441, Santiago, Chile }

\author{ Emanuel Gallo\orcidicon{0000-0002-8974-5134}$^1$, $^2$}
\email{egallo._.at._.unc.edu.ar}
\affiliation{$^1$ Universidad Nacional de Córdoba, Facultad de Matemática, Astronomía, Física y Computación, Grupo de Relatividad y Gravitación; Córdoba, Argentina.\\ $^2$ Consejo Nacional de Investigaciones Científicas y Técnicas,\\ CONICET, IFEG. Córdoba, Argentina. }

\begin{abstract}
In principle, global properties of solution of Einstein equations need to be addressed using the conformal Einstein equations, because  this conformal compactification allows a clean definition of the `infinities' (spacelike, timelike and null infinity) of General Relativity. 
However, in numerical calculations often compactified coordinates in the physical space are used to reach these infinities. 
In this note, we discuss the conformal Einstein equations in spherical symmetry coupled to a massless scalar field and compare them with corresponding equations using a compactified coordinate in physical spacetime. 
The derivation of the field equations is based on metrics, in which the radial coordinate is an affine parameter along outgoing null rays.
We show that the conformal equations within an affine-null metric formulation can be cast in a natural hierarchical form after the introduction of suitable auxiliary fields. 
The system of partial differential equations associated with the resulting (unphysical) conformal field equations proves to be identical to a system that employs a compactified coordinate in physical space 
along with well-constructed regularized fields.  
The reason for this equivalence is the introduction of new regularized fields in the physical spacetime after coordinate compactification to obtain a regular system of equations on the complete domain of the compactified coordinate. 
As part of this work, we also present the solution of the conformal field equations
in affine-null coordinates near the conformal boundary, where the Bondi mass loss formula for a massless scalar field is recovered. 
The validity of the balance law of the mass loss at null infinity is demonstrated numerically.
\end{abstract}

\maketitle
\section{Introduction}
The concept of conformal compactification by Penrose \cite{Penrose:1962ij,Penrose:1964ge} has led to a thorough understanding of the structure and radiative properties of the spacetime of General Relativity. 
It gives mathematical meaning  to paradoxical phrases like `points at infinity', `close to infinity', `a coordinate patch at infinity' or alike.  
The meaning is achieved by adding a boundary $\mathcal{I}$ to the spacetime manifold $\tilde{\mathcal{M}}$, which contains the start and endpoints of all null geodesics that are in $\tilde{\mathcal{M}}$. 
The union of the spacetime manifold and this boundary surface gives rise to a  bigger unphysical manifold $\mathcal{M}$ and it is accomplished by introducing a sufficiently differentiable  non-negative scalar function  $\Omega$  in the unphysical manifold, while conformally relating the metric of the unphysical spacetime $g_{ab}$ with the physical metric $\tilde g_{ab}$, like $g_{ab} = \Omega^2\tilde g_{ab}.$ 
The triple $(\mathcal{M}, g_{ab}, \Omega)$ is the conformally compactified spacetime of the physical spacetime $(\tilde{\mathcal{M}}, \tilde g_{ab})$.
From the conformal metric relation, it is seen immediately that null vectors in the physical spacetime are also null vectors in the unphysical spacetime.
Regarding the previous paradoxical statements, `points at infinity' are now points on the boundary, while points `near infinity'  are in a half-open patch of the unphysical conformal spacetime, which includes the boundary and points not on the boundary. 
{In addition}, sufficient differentiability assumptions need to be made in order to evaluate geometric objects like connection and curvature  and to ensure a connectivity between  boundary and the physical spacetime. 
The mathematical aspects of this `linking' process between the conformal boundary and physical spacetime was studied in great detail by numerous authors, see e.g. review by \cite{Geroch:1977big}. Indeed,  the relationship between physical quantities at finite distances in  vacuum spacetimes and their connection to invariant quantities defined in  conformal spacetimes was studied for applications in numerical relativity by \cite{Lehner:2007ip,Gallo:2008sk,Zenginoglu:2009ey,Helfer:2009iz}.

With the infinite spacetime boundary at hand,  asymptotic properties of radiation fields, in particular gravitational radiation,  could be elegantly  and geometrically understood, \cite{Penrose:1965am,Tamburino:1966zz,Hawking:1968qt}. 
The fact that the conformal spacetime is larger than the physical spacetime as well as that various massless fields have  conformally invariant  field equations \cite{Penrose:1965am} has spawned a wide interest in in studying quantum aspects \cite{Frolov:1977bp} of asymptotic gravitational fields using conformal methods \cite{Ashtekar:1981sf,Ashtekar:1981bq}. 
Not to mention, twistor geometry \cite{Penrose:1967wn,Penrose:1968me,Penrose:1972ia} or the famous AdS/CFT correspondence \cite{Maldacena:1997re,Aharony:1999ti}, who are  candidates for finding a quantum theory of gravity, involve to some extend conformal treatments of the field equations. 

Remaining on the classical side of studies, conformal methods play a role in the understanding of  the nature of the singularities arising from the solution of Einstein equations \cite{Geroch:1968us,Geroch:1972un}. 
Moreover, the attachment of the boundary to the physical spacetime allows a bonafide investigation of the regularity of Einstein's equations and their solutions using the conformal Einstein equations \cite{Friedrich:1981at,Friedrich:1981wx,Friedrich:1986qfi,Friedrich:1991nn,Friedrich:1995vb}.
Based on those works, numerical relativists \cite{Frauendiener:1997zc,Frauendiener:1997ze,Frauendiener:1998yi,Hubner:1999th} used the conformal Einstein equations to find  numerical solutions to these highly non-linear  partial differential equations; the articles \cite{Frauendiener:2002mm,Frauendiener:2000mk,Zenginoglu:2007it} collect  contributions. 
These numerical methods often use the hyperboloidal slicing \cite{Hubner:2000zn,Zenginoglu:2008pw,Rinne:2009qx}, which are spacelike hypersurfaces that become null in the  limit towards the conformal boundary \cite{Beyer:2017qjo}. 
A recent software package to solve the conformal Einstein equations employing  state-of-the-art numerical methods is the COFFEE package \cite{Doulis:2019ejj} and it was applied to find numerical solutions at both future and past null infinity \cite{Frauendiener:2025xcj}.

In a parallel development to these numerical investigations   using the conformal Einstein equations,  the so-called characteristic methods have made astonishing progress in the understanding of radiation fields emanating from isolated sources.
In these characteristic methods, spacetime is foliated by at least one characteristic, i.e. null hypersurface, of the gravitational field.  
There are two powerful formalisms of characteristic methods in General Relativity, the spin-coefficient (or Newman-Penrose) formalism \cite{Newman:1961qr} or the  the metric based approach to of Bondi, Metzner, van der Burg and Sachs \cite{Bondi:1962px,Sachs:1962wk}. 
The former is prominently used for analytic investigations, while the latter has found wide applications in numerical investigations (apart from  analytical breakthroughs like the asymptotic symmetry group, also known as the Bondi-Metzner-Sachs (BMS) group, and the famous Bondi mass loss formula).

Numerical schemes in the Bondi-Sachs formalism base their foundations in the analytical considerations of  \cite[also see \cite{Handmer:2015dsa}]{Tamburino:1966zz} for vaccum spacetimes, who have shown that there is a patch containing null infinity in a conformal spacetime, in which exists a Bondi mass loss formula. 
The first proper numerical investigations in a Bondi-Sachs formalism were done in the early 1980ies by \cite{Isaacson:1983xc}, who compactified the area distance coordinate of the Bondi metric to map null infinity on the numerical grid. 
This coordinate compactification  is now the `standard way' to reach null infinity in characteristic codes employing versions of the Bondi-Sachs metric.  
Various milestones and achievements have been reached with  numerical characteristic Bondi-Sachs codes, for example, Cauchy Characteristic extraction of gravitational waves \cite{Bishop:1996gt}, the first long term stable evolution of a perturbed black hole  \cite{1998PhRvL..80.3915G},  core collapse of relativisitic stars \cite{Linke:2001mq,Siebel:2001ii}, demonstration of  the Choptuik critical phenomena \cite{Choptuik:1992jv, Gundlach:2007gc,Gundlach:2024eds} at null infinity \cite{Purrer:2004nq},
and Cauchy-Characteristic matching
\cite{Ma:2023qjn}; for reviews consult \cite{Winicour:2012znc,Madler:2016xju} or  for a listing of most recent works \cite{Ma:2023qjn}.

An alternative way to the classical Bondi-Sachs metric (or versions thereof) for developing a metric based numerical algorithm is to use an affine-null metric \cite{Win2013}.
The difference to the Bondi-Sachs metric is that the coordinate along the null generators is an affine parameter instead of an area distance. 
The affine parameter coordinate has the advantage that it remains regular at an apparent horizon, where the expansion rate of outgoing null rays vanishes and the area distance coordinate renders the volume element of its defining metric infinite.
Using an affine parameter coordinate in a null metric is beneficial when simulating a forming black hole or in situations of near horizon formation, where the area distance leads to numerical inaccuracies due to its singular behavior \cite{Husa:2001pk}.
However, the price to pay with an affine-null metric is that the nice and convenient Bondi-Sachs hierarchy of its resulting partial differential equations is broken due to the presence of an additional time derivative in one of the equations.
Nevertheless, a hierarchical system of hypersurface-evolution equations can be restored via some rather non-trivial variable transformations \cite{Win2013,Madler:2018bmu,Crespo:2019mcv,Madler:2025ibn}.
Recent numerical investigations \cite{Madler:2024kks} in physical space, also employed a compactified version of the affine parameter to reach null infinity and also extracted the gravitational redshift and Bondi mass decay of supercritical solutions forming a black hole. 
In this work we show how these extracted regularized quantities (including the scalar field) can naturally be obtained from well defined fields at null infinity in the sense of a conformal compactification.
We calculate the conformal field equations in affine-null coordinate in a conformally compactified spherically symmetric spacetime filled with a massless scalar field.
More importantly, we will demonstrate how to successfully obtain a hierarchical set of equations in the conformal metric, utilizing field variables that are regular and well-defined throughout the entire non-physical spacetime, including its future null boundary.
The obtained conformal field equations are then compared with those resulting from a coordinate transformation in physical space. 
During the course of this work we also derive a solution of the affine-null metric equations in the neighborhood of the conformal boundary $\mathcal{I}$, which we have not seen in this context. 

In Sec.~\ref{sec:conf}, we present the conformal field equations in hierachical form and give their solution at $\mathcal{I}$.
For comparison, in Sec.~\ref{sec:compact}, we present the hypersurface-evolution equations with a compactified physical affine parameter, while no conformal boundary is introduced.
In Sec.~\ref{sec:num}, we apply our results to extract the scalar news function at null infinity.
Our results are discussed 
in Sec.~\ref{sec:discuss}.
In App.~\ref{sec:appendix}, we repeat for completeness the analysis of Sec.~\ref{sec:conf} and \ref{sec:compact} using the classical Bondi-Sachs metric. 
In App.~\ref{App:sch}, we derive  a general  representation of the Schwarzschild metric in conformal  affine-null formulation  serving for comparison with its Bondi (inertial) counterpart.

We use  units with $G=1$, $c=1$  and label the gravitational constant with $\kappa$. The metric signature is $+2$ and we denote with a tilde coordinates and tensors (like metric, scalar field, Ricci tensor, covariant derivative, etc) in physical space, while those quantities without a tilde are in conformal space. This nomenclature is the one of \cite{Penrose:1964ge} and \cite{Geroch:1977big}.
%%%%%%%%%%%%%%%%%%%%%%%%%%%%%%%%%%%%%%%%
\section{Conformal compactification and field equations}\label{sec:conf}

We consider a four dimensional smooth physical spacetime $(\tilde {\mathcal{M}}, \tilde g_{ab})$ filled with a massless physical scalar field $\tilde \Phi$. 
The spacetime is charted with  associated coordinates $\tilde x^a$  and the Levi-Civita covariant derivative with respect to $\tilde g_{ab}$ is $\tilde \nabla_a$.  
The field equations are the Einstein scalar field equations 
\begin{equation}\label{eq:FE_phys_}
    \tilde R_{ab} = \kappa (\tilde \nabla_a\tilde \Phi)(\tilde \nabla_b\tilde \Phi)\;\;,\;\;
    0=\tilde \nabla^a\tilde \nabla_a\tilde \Phi,
\end{equation}
where $\kappa$ is the gravitational  constant. 
In this physical spacetime exists an invariant concept of mass, the Misner-Sharp mass, defined via \cite{Misner:1964je},
\begin{equation}\label{eq:MisnerSharp_phys}
\tilde M = \frac{r}{2}\left[1-\tilde g^{ab}(\tilde \nabla_a r)(\tilde \nabla_b r)\right],
\end{equation}
where $r$ is an area distance. Its asymptotic limit relates to the Bondi mass $m_B=\lim_{\tilde \lambda\rightarrow\infty}\tilde M$, where $\tilde \lambda$ is an affine parameter along outgoing null rays.

Let $\mathcal{I}$ be the manifold containing the start and endpoints of all null geodesics in $\tilde{\mathcal{M}}$.
The unphysical manifold ${\mathcal{M}} =\tilde{ \mathcal{M}}\cup \mathcal{I}$ is the closure of $\tilde{\mathcal{M}}$ with a three dimensional boundary i.e. $\mathcal{I} = \partial \tilde{\mathcal{M}}$.
In $\mathcal{M}$ exists an additional smooth scalar field $\Omega$ with the following properties: 
i) it is positive in $\mathcal{M}/\mathcal{I}$,  
ii) vanishes on $\mathcal{I}$, and iii) its gradient is non vanishing  on $\mathcal{I}$ \cite{Geroch:1977big,Griffiths:2009dfa}.
 
To relate the  physical spacetime $(\tilde{\mathcal{M}}, \tilde g_{ab})$ and the unphysical spacetime, % $(\tilde{\mathcal{M}}, \tilde g_{ab})$, 
the corresponding unphysical metric $g_{ab}$ is defined via the conformal relation
\begin{equation}\label{eq:conf_phys_g}
g_{ab} = \Omega^2\tilde g_{ab},
\end{equation}
while the unphysical scalar field $\Phi$ in $\mathcal{M}$ is related with the physical one $\tilde \Phi$ via
\begin{equation}\label{eq:conf_Phi}
    \Phi = \frac{\tilde \Phi}{\Omega}.
\end{equation}
We remark that the conformal factor $\Omega$ is up to its defining properties arbitrary in this construction.

Since $\Omega=0$ and $\nabla_a\Omega\neq0$ on $\mathcal{I}$, a finite value  of $\Phi$ on $\mathcal{I}$  requires    that $\tilde \Phi$ is sufficiently differentiable in a coordinate patch near $\mathcal{I}$ and that $\tilde \Phi$ vanishes at in the limit towards $\mathcal{I}$. In other words, the physical field $\tilde \Phi$ requires a local smooth extension from $\mathcal{M}/\mathcal{I}$ to $\mathcal{I}$ \cite{Geroch:1977big}.
Unphysical coordinates in $\mathcal{M}$ are denoted with $x^a$ and the Levi-Civita covariant derivative with respect to $g_{ab}$ is $\nabla_a$. 

The Ricci tensor of the physical and unphysical spacetimes are related like
\begin{align}\label{eq:conf_ricci}
\tilde R_{ab}(\tilde g_{cd}) =&   R_{ab}( g_{cd})+ \frac{2  \nabla_a \nabla_b \Omega}{\Omega}\nonumber
\\&
     +   g_{ab}\Big[\frac{ \nabla^c  \nabla_c \Omega}{\Omega} -\frac{3( \nabla^c\Omega )(\nabla_c\Omega)}{\Omega^2}\Big].
\end{align}
Using the field equations \eqref{eq:FE_phys_} in physical space we find the field equations for the unphysical spacetime $(\mathcal{M}, g_{ab}, \Omega)$
\begin{subequations}\label{eq:FE_conf_an}
\begin{align}
0=&-\kappa \Big( \nabla_a \Omega \Phi\Big)\Big( \nabla_b \Omega \Phi\Big)
+   R_{ab}( g_{cd})+ \frac{2  \nabla_a  \nabla_b \Omega}{\Omega}\nonumber
\\
&
     +   g_{ab}\Big[\frac{ \nabla^c  \nabla_c \Omega}{\Omega} -\frac{3( \nabla^c\Omega )( \nabla_c\Omega)}{\Omega^2}\Big],\\
0=&\Omega^3 \nabla_a  \nabla^a\Phi
+ \Omega^2\Phi  \nabla_a  \nabla^a\Omega
-2\Omega\Phi  g^{ a b}   \Omega_{, b}  \Omega_{, a}.
\end{align}
\end{subequations}
It can easily be shown that the Misner sharp mass \eqref{eq:MisnerSharp_phys} is not conformally invariant because the physical area distance $r$ transforms under    conformal rescaling 
\eqref{eq:conf_phys_g} of the metric  to  a conformal (unphysical) area distance like $\mathcal{R}: = \Omega r$. 
Therefore, the Misner sharp mass behaves like 
\begin{equation}\label{eq:Misner_Sharp_conf}
 \tilde M = \frac{M}{\Omega}+\frac{\mathcal{R}^2}{\Omega^2}\left[ (\nabla^a\Omega)(\nabla_a\mathcal{R}) -\frac{\mathcal{R}}{2\Omega} (\nabla^b\Omega)(\nabla_b\Omega)\right]  ,
\end{equation}
where $M$ is the conformal (unphysical) Misner-Sharp mass
\begin{equation}\label{eq:conformal_MS}
M = \frac{\mathcal{R}}{2}\left[1-g^{ab}(\nabla_a \mathcal{R})(\nabla_b \mathcal{R})\right]
\end{equation}
determined by an area distance $\mathcal{R}$ in the unphysical conformal spacetime $(\mathcal{M}, g_{ab}, \Omega)$.

%%%%%%%%%%%%%%%%%%%%%%%%%%%%%%%%%%%%%%%%%%%%%%%%%%%%%%%%%%%%%%%%%%%%
\subsection{Conformal compactification with affine-null coordinates in spherical symmetry}
To establish a particular frame for the field equations in unphysical space,  we begin with constructing a metric in the  physical spacetime.

Consider a spherically symmetric four dimensional physical smooth manifold $\tilde{\mathcal{M}}$ charted with affine-null coordinates $\tilde x^a = (\tilde u,\tilde \lambda, \tilde x^A)$. 

The $\tilde u$ coordinate labels a family of outgoing null cones $\tilde u=const$.
The surface forming null rays in the $\tilde u=const$ hypersurfaces are labelled with angular coordinates $\tilde x^A = (\tilde x^2, \tilde x^3)$. 
Therefore only the coordinate $\tilde x^1=\tilde \lambda$  varies along the generators of the null rays in the $\tilde u=const$ hypersurfaces.
The coordinate $\tilde \lambda$ is an affine parameter defined via the relation $(\tilde\nabla^a \tilde u)(\tilde \nabla_a \tilde \lambda) = -1$. 
The area of the two spaces $\tilde \lambda=const$ and $\tilde u=const$ is taken to be $4\pi r^2$, where $r$ is an area distance.
The intrinsic metric of these spaces has the conformal representation $r^2(\tilde u, \tilde \lambda)q_{AB}(\tilde x^C)$  with $q_{AB}$ being a unit sphere metric with respect to $\tilde x^A$.
If the null cones have vertices where $r=\tilde\lambda=0$, those vertices trace the geodesic world line of a freely falling observer along the central geodesic of spherical symmetry and  regularity conditions need to be applied \cite{Manasse:1963zz,Poisson:2009pwt,Madler:2012sg}. 

The spherically symmetric affine-null metric  has the line element \cite{Husa:2001pk,Win2013,Crespo:2019mcv,Madler:2018bmu,Madler:2023aqc,Madler:2024kks,Madler:2025ibn}
\begin{equation}\label{eq:AN_metric_phys}
d\tilde s_{AN}^2 = -W d\tilde u ^2 -2d\tilde u d\tilde \lambda +  r^2 q_{AB}d\tilde x^A d\tilde x^B
\end{equation}
and its nontrivial covariant components and volume element are
\begin{equation}
\begin{split}
&\tilde g^{\tilde u \tilde \lambda } = -1\;\;,\;\;
\tilde g^{ \tilde\lambda  \tilde\lambda } = W\;\;,\;\;
\tilde g^{\tilde A \tilde B } = \frac{q^{\tilde A \tilde B }}{ r^2} \;\;,\;\;\\
&\sqrt{-\det(\tilde g_{ab})} =  r^2 \sqrt{\mathfrak{q}}   
\end{split}
\end{equation}
with $\mathfrak{q}=\det(q_{AB})$.

The scalar field obeys the Sommerfeld radiation condition, meaning it vanishes at large distances in the limit taken along outgoing null rays, i.e. 
 \begin{equation}\label{eq:SommerFeld}
     \lim_{\tilde \lambda\rightarrow \infty\atop \tilde u=const}\tilde\lambda\tilde \Phi<\infty.
 \end{equation}

Since \eqref{eq:conf_phys_g} can be expressed in physical space coordinates $\tilde x^a$ as well as in conformal coordinates $ x^a$, due to the requirements of the smoothness of the conformal factor and conformal metric, we propose a conformal factor in physical space coordinates, which  
is  subsequently transformed to provide the definition of the conformal  coordinates. 
In  coordinates $\tilde x^a$ of the affine-null metric \eqref{eq:AN_metric_phys}, $\tilde\lambda$ is non-negative and extend to positive infinity. The  conformal factor 
\begin{equation}\label{eq:proposes_omega}
\Omega(\tilde x^a) := \Omega(\tilde \lambda) = \frac{1}{1+\tilde\lambda}
\end{equation}
is  positive  for every value value of $\tilde \lambda$ in $\tilde{\mathcal{M}}$ and it vanishes in the limit $\tilde \lambda\rightarrow\infty$. Then using \eqref{eq:proposes_omega}, the calculation of $g_{ab}$ in physical coordinates results in
\begin{widetext}
\begin{align}
ds^2_{AN}=\Omega^2 \tilde g_{ab}d\tilde{x}^ad\tilde{x}^b 
= &
-\frac{W(\tilde u , \tilde \lambda)}{(1+\tilde\lambda)^2}d\tilde u^2 -\frac{2d\tilde u d\tilde \lambda}{(1+\tilde\lambda)^2} + \left[\frac{r(\tilde u,\tilde\lambda)}{1+\tilde\lambda}\right]^2q_{AB}d\tilde x^Ad\tilde x^B\;\;,\\
= &
-\frac{W(\tilde u , \tilde \lambda)}{(1+\tilde\lambda)^2}d\tilde u^2 -2d\tilde ud\left(\frac{\tilde\lambda}{1+\tilde\lambda}\right) 
+ \left[\frac{r(\tilde u, \tilde \lambda)}{1+\tilde\lambda}\right]^2q_{AB}d\tilde x^Ad\tilde x^B\;\;,
\end{align}
\end{widetext}
which allows us to define the conformal  coordinates
\begin{equation}\label{def:conf_ccords}
x^a:=(u, x, x^A) = \left(\tilde u, \frac{\tilde \lambda}{1+   \tilde\lambda}, \tilde x^A\right)\;\;.
\end{equation}
where $x\in[0,1]$.
Consequently, we have
\begin{equation}\label{eq:tildelambdax}
\tilde \lambda(x) = \frac{x}{1-x}
\end{equation}
and  the conformal factor \eqref{eq:proposes_omega} becomes
\begin{equation}\label{eq:omega_def}
\Omega(x^a) = 1-x\;\;.
\end{equation}
In conformal coordinates $x^a$, $\tilde \lambda(0)=0$ and the asymptotic limit of $\tilde \lambda$ corresponds to $x=1$.
The conformal factor \eqref{eq:omega_def} vanishes if $x=1$ and its gradient $\nabla_a \Omega$ is finite and nonzero there. 
The line element of the unphysical metric $g_{ab}$ in $x^a$ coordinates is 
\begin{equation}\label{eq:metric_an_conf}
ds^2 = -(1-x)\mathcal{W}du^2-2dudx + \mathcal{R}^2q_{AB}dx^Adx^B\;,
\end{equation}
where we defined the naturally occurring  variables
$
\mathcal{R} := (1-x)r$ and $\mathcal{W} := (1-x)W$.
Also note that $x$ in \eqref{eq:metric_an_conf} is an affine parameter since $g_{ux}=-1$.
The metric \eqref{eq:metric_an_conf} at $x=1$ takes  the form
\begin{equation}
ds^2 \Big|_{x=1}= 0\cdot du^2  +\mathcal{R}^2q_{AB}dx^Adx^B \Big|_{x=1}\;,
\end{equation}
showing that $x=1$ is a null hypersurface, because the metric degenerates to a three-dimensional null metric.
Given \eqref{eq:omega_def}, we find the relation between the physical scalar field $\tilde\Phi$ and the conformal scalar field as $\Phi$ as ${\tilde \Phi = (1-x)\Phi}$. 

This allows us to write the relevant conformal field equations \eqref{eq:FE_conf_an} for \eqref{eq:metric_an_conf} using  \eqref{eq:omega_def} to find
\begin{widetext}
\begin{subequations}
\begin{eqnarray}
0&=&\mathcal{R}_{,uu}+
\frac{1}{2}(1-x)^2\left[\left(\frac{\mathcal{W}}{1-x}\right)_{,x}\mathcal{R}_{,u}-\left(\frac{\mathcal{R}}{1-x}\right)_{,y}\mathcal{W}_{,u}
 \right]
 %\nonumber\\
 %&&
 -\frac{1}{4}(1-x)^3\mathcal{W}\left[\mathcal{R}\left(\frac{\mathcal{W}}{1-x}\right)_{,x}\right]_{,x}
+\frac{\kappa}{2}\mathcal{R}[(1-x)\Phi_{,u}]^2
,\nonumber\\
&&
\label{eq:supp_conf_an}\\
0&=&\mathcal{R}_{,xx}+\frac{\kappa}{2} \mathcal{R}\Big[(1-x)\Phi_{,x}-\Phi\Big]^2,\label{eq:main_conf_an_R11}\\
0&=& q_{AB}\left\{-\frac{1}{2}\left[ (1-x)(\mathcal{R}^2)_{,x}\mathcal{W} - 2x - 2(\mathcal{R}^2)_{,u}\right]_{,x}  
%\right.\nonumber\\
%&&\left.
- \frac{(1-x)^2}{\mathcal{R}^2}\left[\frac{\mathcal{R}^4\mathcal{W}}{(1-x)^2}\right]_{,x}
+\frac{2(\mathcal{R}^2)_{,u}}{1-x}\right\},\label{eq:FE_tr_an_rW_renorm}\label{eq:main_conf_an_RAB}\\
0&=&-(\mathcal{R}^2\Phi_{,u})_{,x}-(\mathcal{R}^2\Phi_{,x})_{,u}+\left[\mathcal{R}^2(1-x)\mathcal{W}\Phi_{,x}\right]_{,x}
%\nonumber\\
%&&
+\left\{\frac{(\mathcal{R}^2)_{,u}}{(1-x)} - (1-x)\left[\frac{\mathcal{R}^2\mathcal{W}}{1-x}\right]_{,x}\right\}\Phi.\label{eq:main_conf_an_wave}
\end{eqnarray}
\end{subequations}
\end{widetext}

According to the twice contracted Bianchi identities and the local conservation of energy-momentum the first equation, \eqref{eq:supp_conf_an}, is the supplementary equation. In a numerical algorithm, it is used to fix the boundary conditions of the fields. The equations \eqref{eq:main_conf_an_R11}-\eqref{eq:main_conf_an_wave} are the main equations used to evolve the fields. 
But due to the presence of the $\mathcal{R}_{,u}$ derivative, equations \eqref{eq:main_conf_an_R11}-\eqref{eq:main_conf_an_wave} are not suited for numerical work, since they do have not the typical hierarchy due to the choice of a characteristic foliation like e.g. in the Bondi-Sachs case \cite{Bondi:1962px,Sachs:1962wk,Tamburino:1966zz,Madler:2016xju}.

In order to restore a hierarchy, we start by introducing  the  field
\begin{eqnarray}
    \mathcal{Y} &=&  (1-x)\left(\frac{\mathcal{R}}{1-x}\right)_{,x}\mathcal{W}-\frac{2\mathcal{R}_{,u}}{1-x} , 
\end{eqnarray}
which casts \eqref{eq:FE_tr_an_rW_renorm} into 
\begin{equation}\label{eq:ode_Y_conf}
    0=\left(\frac{\mathcal{RY}_{,x}}{1-x} - \frac{x}{1-x}\right)_{,x}.
\end{equation}
Eqn. \eqref{eq:ode_Y_conf} has the first integral
\begin{equation}\label{eq:sol_y_conf}
\mathcal{Y} = \frac{x+(1-x)\mathcal{Z}_0(u)}{\mathcal{R}},
\end{equation}
with  $\mathcal{Z}_0(u)$ being a function of integration. Next, definition of 
\begin{equation}\label{eq:def_L}
\begin{split}
\mathcal{L} = &2\mathcal{R}\Phi_{,u}-\frac{[2\mathcal{R}_{,u}-(1-x)\mathcal{Y}] [(1-x)\Phi]_{,x}}{(1-x)\mathcal{R}_{,x}+\mathcal{R}}
\\
=&2\mathcal{R}\Phi_{,u}-\frac{\mathcal{W}[(1-x)\Phi]_{,x}}{(1-x)\mathcal{R}_{,x}+\mathcal{R}}   \end{split}
\end{equation}
gives a first order evolution equation for the unphysical scalar field $\Phi$, i.e.
\begin{equation}
    \Phi_{,u}
=\frac{1}{2}\left\{\frac{\mathcal{L} }{\mathcal{R}} +\mathcal{W}[(1-x) \Phi]_{,x}\right\},
\end{equation}
and casts \eqref{eq:main_conf_an_wave} into a hypersurface equation
\begin{equation}
    \mathcal{L} _{,x} = 
\mathcal{Y}[(1-x)\Phi]_{,x}
=\frac{[x+(1-x)\mathcal{Z}_0][(1-x) \Phi]_{,x}}{\mathcal{R}} .
\end{equation}
What remains, is to find a hypersurface equation for the variable $\mathcal{W}$.  After a  lengthy and tedious calculation using \eqref{eq:main_conf_an_R11}
 we arrive at
\begin{align}
\mathcal{W}_{,xx}=&
     \kappa\frac{[(1-x)\Phi]_{,x} \mathcal{L}}{\mathcal{R}} +\frac{2[\mathcal{R}+(1-x)\mathcal{R}_{,x}]\mathcal{Z}_0}{\mathcal{R}^3}
     \nonumber\\
     &
     -\frac{1}{x}\left(\frac{x^2}{\mathcal{R}^2}\right)_{,x}.
\end{align}
In summary,  we are left with the hierarchical system of partial differential equations 
\begin{subequations}\label{eq:hierachy_conf}
    \begin{align}
0=&\mathcal{R}_{,xx} + \frac{\kappa \mathcal{R}}{2}  \{[(1-x) \Phi]_{,x}\}^2,\\
\mathcal{L} _{,x} =& 
\frac{[x+(1-x)\mathcal{Z}_0][(1-x)\Phi]_{,x}}{\mathcal{R}},\\
%\frac{\mathcal{Z}[(1-x) \Phi]_{,x}}{\mathcal{R}} \\
%
\mathcal{W}_{,xx}=&
     \kappa\frac{[(1-x)\Phi]_{,x} \mathcal{L}}{\mathcal{R}} +\frac{2[\mathcal{R}+(1-x)\mathcal{R}_{,x}]\mathcal{Z}_0}{\mathcal{R}^3}\nonumber\\
     &
     -\frac{1}{x}\left(\frac{x^2}{\mathcal{R}^2}\right)_{,x},
     \\
\Phi_{,u}
=&\frac{1}{2}\left\{\frac{\mathcal{L} }{\mathcal{R}} +\mathcal{W}[(1-x) \Phi]_{,x}\right\},
\end{align}
\end{subequations}
which is regular at the boundary $\mathcal{I}$, i.e. it has no singularities at $x=1$. 
The system \eqref{eq:hierachy_conf} is completed   with \eqref{eq:supp_conf_an} and \eqref{eq:main_conf_an_RAB} evaluated a some value $x\in[0,1)$, which are needed to fix the boundary values for the hypersurface fields $\mathcal{R}$, $\mathcal{Z}$, $\mathcal{L}$ and $\mathcal{W}$.
%%%%%%%%%%%%%%%%%%%%%%%%%%%%%%%%%%%%%%%%%%%%%%%
\subsection{Regularity condition of the fields at the vertex of the null cone and the respective field equations}

If the system \eqref{eq:hierachy_conf} is solved for $x\ge0$ and $\mathcal{R}\ge 0$, it still has singularities at the conformal areal distance $\mathcal{R}=0$. 
Then $\mathcal{R}=0$ is a vertex of a null cone and the coordinate singularity is due to the singular nature of the vertex. It can then be dealt with the requirements of the Fermi normal coordinate system (imposed in the physical spacetime, see e.g. \cite{Poisson:2011nh,Madler:2012sg}) at $x=0$, which imply
\begin{equation}\label{eq:reg_cond_conf}
\begin{split}
        \mathcal{R}(u,0) &= 0\;\;,\;\;\mathcal{R}_{,x}(u,0)=1\\
    \mathcal{W}(u,0) &= 1\;\;,\;\;\mathcal{W}_{,x}(u,0)=-1.
    \end{split}
\end{equation}
Using the \eqref{eq:reg_cond_conf}, \eqref{eq:sol_y_conf} and \eqref{eq:def_L}, these regularity conditions imply $\mathcal{Z}_0=0$ and $\mathcal{L}(u,0)=0$.
Assuming a regular Taylor series expansion of the scalar field at the vertex like
\begin{equation}\label{eq:phi_vertex}
\Phi(u,x) = \Phi_0(u)+\Phi_{1}(u)x + O(x^2),
\end{equation}
the right hand side of \eqref{eq:hierachy_conf} are given by the regular expressions
\begin{align}
\mathcal{R}_{,xx}(u,0)=&0,\\
\mathcal{L}_{,x}(u,0)=&\Phi_{1}(u),\\
\mathcal{W}_{,xx}(u,0)=&\kappa \Phi_1^2(u),\\
\Phi_{,u}(u,0)=&\Phi_{1}(u).\label{eq:dphidu_vertex}
\end{align}
Finally, after using these regularity conditions,  \eqref{eq:hierachy_conf} further simplifies to
\begin{subequations}\label{eq:hierachy_conf_final}
    \begin{align}
0=&\mathcal{R}_{,xx} + \frac{\kappa \mathcal{R}}{2}  \{[(1-x) \Phi]_{,x}\}^2,\label{eq:hierachy_conf_final_hyp1}\\
\mathcal{L} _{,x} =& 
\frac{x[(1-x)\Phi]_{,x}}{\mathcal{R}},\label{eq:hierachy_conf_final_hyp2}\\
%\frac{\mathcal{Z}[(1-x) \Phi]_{,x}}{\mathcal{R}} \\
%
\mathcal{W}_{,xx}=&
     \kappa\frac{[(1-x)\Phi]_{,x} \mathcal{L}}{\mathcal{R}} -\frac{1}{x}\left(\frac{x^2}{\mathcal{R}^2}\right)_{,x}
     \nonumber\\
     =&
     \kappa\frac{  (\mathcal{L}^2)_{,x}}{2x} 
     -\frac{1}{x}\left(\frac{x^2}{\mathcal{R}^2}\right)_{,x}
     \label{eq:hierachy_conf_final_hyp3}\\
\Phi_{,u}
=&\frac{1}{2}\left\{\frac{\mathcal{L} }{\mathcal{R}} +\mathcal{W}[(1-x) \Phi]_{,x}\right\}.\label{eq:hierachy_conf_final_ev}
\end{align}
\end{subequations}
For the system \eqref{eq:hierachy_conf_final_ev}, the regularity conditions also assure that supplementary equation \eqref{eq:supp_conf_an} holds automatically.
%%%%%%%%%%%%%%%%%%%%%%%%%%%%%%%%%%%%%%%%%
\subsection{Solution of the field equations at the conformal boundary }\label{sec:Hreds}

At the boundary $\mathcal{I}$, i.e. at $x=1$, we assume that the conformal scalar field has the regular expansion
\begin{equation}\label{eq:expansion_phi_I}
    \Phi(u,x) = \Phi_{[0]}(u)+\Phi_{[1]}(u)(1-x) + O[(1-x)^2],
\end{equation}
which we can also write as
\begin{equation}\label{eq:expansion_phi_I}
    \Phi(u,x) = \Phi_{[0]}(u)+\Phi_{[1]}(u)\Omega(x) + O(\Omega^2).
\end{equation}
Hereafter, this will abbreviate the writing of most of the expressions. 

In terms of the physical scalar field $\tilde \Phi$ , this expansion corresponds to 
\begin{equation}\label{eq:expansion_phi_I_phys_ux}
    \tilde\Phi(u,x) = (1-x)\Phi_{[0]}(u)+\Phi_{[1]}(u)(1-x)^2 + O[(1-x)^3].
\end{equation}
or in terms of the coordinates  $\tilde x^a$ using \eqref{def:conf_ccords}
\begin{equation}\label{eq:expansion_phi_I_phys_ulambda}
    \tilde\Phi(\tilde u,\tilde \lambda) = \frac{\Phi_{[0]}(u)}{\tilde \lambda}+\frac{\Phi_{[1]}(u)}{\tilde \lambda^2} + O(\tilde \lambda^{-3}),
\end{equation}
which shows that $\Phi_{[0]}$ is related to the monopole of the scalar field.

{To find the most general solution of the conformal field equations near $\mathcal{I}$, we insert
\eqref{eq:expansion_phi_I} into \eqref{eq:hierachy_conf}. 
 
Subsequent integration of the respective hypersurface equations  yields the following conformal space solution}
\begin{subequations}\label{eq:sol_conf_general}
\begin{eqnarray}
    \mathcal{R}(u,x)&=&H + \mathcal{R}_{[1]}\Omega-{\frac{\kappa H\Phi_{[0]}^2}{4} \Omega^2}
    +O[\Omega^3]\\
    \mathcal{Z}(u,x) &=& \mathcal{Z}_0(u)\\
    \mathcal{L}(u,x)&=&\mathcal{L}_{[0]} + \frac{\Phi_{[1]}}{H}\Omega+\nonumber\\&&
    +\left[\frac{\Phi_{[0]}(\mathcal{Z}_0-1)+2\Phi_{[1]}}{2H}-\frac{\Phi_{[0]}\mathcal{R}_{[1]}}{2H^2}\right]\Omega^2\nonumber\\&&+O[\Omega^3]\\
    \mathcal{W}(u,x)&=&\mathcal{W}_{[0]} + \mathcal{W}_{[1]}\Omega
    \nonumber\\&&
    - \frac{  \kappa H^2 \mathcal{L}_0  \Phi_0  + 2(H + \mathcal{R}_{[1]})
    {-2\mathcal{Z}_0H}}{ 2H^3 }\Omega^2
    \nonumber\\&&
    +O[\Omega^3]
\end{eqnarray}
\end{subequations}
%}
where $H$, $\mathcal{R}_{[1]}$, $\mathcal{Z}_0(u)$, $\mathcal{L}_0$, $\mathcal{W}_0$ and $\mathcal{W}_1$ are functions of integration depending on $u$, recall $\mathcal{Z}_0(u)$ arose as function of integration from \eqref{eq:ode_Y_conf} as part of the construction of \eqref{eq:hierachy_conf}. Those functions are not completely independent and only two have physical significance. The dependence of between those coefficients is found by inserting the solutions \eqref{eq:sol_conf_general} into the field equation \eqref{eq:FE_tr_an_rW_renorm}. This operation gives the following relations
\begin{align}
    \mathcal{W}_{[0]}=&\frac{d}{du}\ln H^2,\label{eq:w0}\\
    \mathcal{W}_{[1]}=&\frac{1}{H^2}+\frac{2}{H}\frac{d}{du}\mathcal{R}_{[1]}.\label{eq:w1}
\end{align}
Subsequent insertion of \eqref{eq:sol_conf_general} into the evolution equation \eqref{eq:hierachy_conf_final_ev} while using the expressions for $\mathcal{W}_{[0]}$ and $\mathcal{W}_{[1]}$ determines at leading order the field $\mathcal{L}_{[0]}$, 
\begin{equation}
\mathcal{L}_{[0]}=2\frac{d}{du}(H\Phi_{[0]}).
\end{equation}
Calculation of the leading order of the Misner-Sharp mass \eqref{eq:Misner_Sharp_conf} gives
\begin{equation}
    \tilde M = \frac{1}{2}[H(1-\mathcal{Z}_0)+\mathcal{R}_{[1]}] + O[(1-x)].
\end{equation}
Since the limit $x=1$ corresponds to the limit $\tilde \lambda\rightarrow\infty$ in physical space, we find the Bondi mass as
\begin{equation}\label{eq:Bmass_conf}
m_B = \frac{1}{2}[H(1-\mathcal{Z}_0)+\mathcal{R}_{[1]}] ,
\end{equation}
so that we can replace $\mathcal{R}_{[1]}$ with the Bondi mass.

Inserting \eqref{eq:expansion_phi_I} and the solution \eqref{eq:sol_conf_general} into \eqref{eq:supp_conf_an} gives us at leading orders
a balance equation
\begin{align}\label{eq:mass_loss_HR1dPhi0}
    0= \frac{d}{du}\left[H(1-\mathcal{Z}_0)+\mathcal{R}_{[1]}\right]+ \kappa H \left(\frac{d}{du} H\Phi_{[0]}\right)^2
\end{align}
We can see that the solution of the conformal field equations at the boundary contains three free functions, $H$, $\mathcal{R}_{[1]}$ and $\Phi_{[0]}$, which are coupled via the balance equation \eqref{eq:mass_loss_HR1dPhi0} to one another. 
The quantity 
\begin{equation}\label{eq:monopole}
    C(u) = \lim_{x\rightarrow1}\mathcal{R}\Phi = H\Phi_{[0]},
\end{equation}
corresponds in physical space to the limit $\lim_{r\rightarrow\infty}r\tilde \Phi$, which is the  monopole of the scalar field in physical space.
With the Bondi mass \eqref{eq:Bmass_conf} and the scalar monopole, the balance law \eqref{eq:mass_loss_HR1dPhi0} becomes
\begin{equation}\label{eq:BmassLoss_C}
    H\frac{d}{du}m_B= 
    -\frac{1}{2} \kappa  \left(H\frac{d}{du} C\right)^2\;\;,
\end{equation}
where we introduced an additional factor of $H$, which becomes clear shortly.

Consider the case, where the scalar field is not radiative and consequently $\Phi_{[0]}=0$. 
Then \eqref{eq:BmassLoss_C} implies $m_B=const$ and the metric becomes
\begin{equation}\label{eq:schw_gen_conf}
\begin{split}
    ds^2 =& -\Omega\left[\mathcal{W}_0 +\mathcal{W}_1\Omega -\frac{2m_B}{H^3}\Omega^3  +O(\Omega^4)\right]du^2
     \\&-2dudx 
     +\Big[H 
    +   \mathcal{R}_{[1]} \Omega +O(\Omega^2)\Big]^2q_{AB}dx^Adx^B\;,
\end{split}    
\end{equation}
which corresponds to an asymptotic Schwarzschild solution. 
In particular, if $H=-\mathcal{R}_{[1]} =1$, we recover
the classical conformal version \cite{Penrose:1964ge} of the Schwarzschild solution inertial coordinates (see App. ~\ref{App:sch}).

We note that the general null coordinate $u$ is related to $u_b$ in an inertial frame (also known as Bondi frame) by
\begin{equation}\label{eq:conf_redshift}
    du_b = \frac{du}{H}
    \;\;\Leftrightarrow\;\;
    \frac{d u_b}{d u} = \frac{1}{H},
\end{equation}
where, in particular,  if $H=1$ the frame at $\mathcal{I}$ is called an inertial conformal frame \cite{Tamburino:1966zz, Handmer:2015dsa, Madler:2016xju}. 
The function $H$ can therefore be understood as a redshift factor between an inertial frame and a general frame on $\mathcal{I}$. 
Indeed,  if $H\rightarrow 0$, the conformal coordinate $u$ is at infinite (temporal) distance from an inertial conformal observer, and a signal takes infinite time of propagation between the two frames.   In general $H\rightarrow 0$ implies the formation of an event horizon \cite{Crespo:2019mcv,Madler:2024kks}.  
But, whenever $H\neq0$ on $\mathcal{I}$, we can always transform to an inertial conformal frame with Bondi time $u_b$ where $H=1$ using \eqref{eq:conf_redshift}. Evaluation of \eqref{eq:BmassLoss_C} in an inertial frame gives
\begin{equation}\label{eq:BmassLoss_News}
   0= \frac{d}{du_b}m_B+\frac{1}{2} \kappa N^2(u_b) 
\end{equation}
where we defined the news function 
\begin{equation}\label{eq:def_news}
    N(u_b)=  \frac{dC}{du_b} \;\;.
\end{equation}
The balance law \eqref{eq:BmassLoss_News} is the conformal space analogue of the Bondi mass loss formula for a massless scalar field in physical space, see e.g. \cite{Gomez:1992fk,Madler:2025ibn}. The variables in \eqref{eq:BmassLoss_News} are completely determined by fields {\it on} the boundary $\mathcal{I}$ and in the unphysical spacetime.

%%%%%%%%%%%%%%%%%%%%%%%%%%%%%%%%%%%%%%%%%%%%%%%%%%%%%%%%%%
\section{Coordinate compactification in physical space}\label{sec:compact}
In this section, for comparison, we present the affine-null hierarchical version of the Einstein equations in the physical space with corresponding regularized fields. 

The relevant components of  field equations \eqref{eq:FE_phys_} for the affine-null metric \eqref{eq:AN_metric_phys} in physical space are
\begin{align}
0=&
 r_{,\tilde u\tilde u}   
-\frac{V (r^2V_{,\tilde \lambda})_{,\tilde \lambda} }{4r}
+\frac{1}{2}\left(
 r_{,\tilde u}V_{,\tilde \lambda}  - r_{,\tilde \lambda}V_{,\tilde u} \right)
+\frac{1}{2}\kappa r 
(\tilde \Phi_{,\tilde u})^2 \label{eq:supp_phys},
\\
0=&r_{,\tilde \lambda\tilde \lambda} 
+\frac{\kappa}{2}r(\tilde \Phi_{,\tilde \lambda})^2\label{eq:hyp_phys_r},\\
0=&[Vrr_{,\tilde \lambda}-\tilde \lambda - 2rr_{,\tilde u}]_{,\tilde \lambda}\;,\\
0=&(r^2\tilde \Phi_{,\tilde \lambda})_{,\tilde u}+(r^2\tilde \Phi_{,\tilde u})_{,\tilde \lambda}-(r^2 V\tilde \Phi_{,\tilde \lambda})_{,\tilde \lambda}\label{eq:hyp_phys_wave}.
\end{align}
where \eqref{eq:supp_phys} is the supplementary equation to determine the boundary value evolution and \eqref{eq:hyp_phys_r}-\eqref{eq:hyp_phys_wave} are the main equations for formulating a hypersurface-evolution algorithm.

To cast the main equations into a hierarchy, we introduce the new variables \cite{Madler:2025ibn}
\begin{subequations}
\begin{eqnarray}
Z &= & Vrr_{,\tilde\lambda}  - 2 rr_{,\tilde u} - \tilde \lambda\;\;,\;\;\\
\label{eq:defL}
{L} &=& \frac{2r\left(  r_{,\tilde\lambda}\Phi_{,\tilde u}-r_{,\tilde u}\Phi_{,\tilde \lambda}\right)- (Z+\tilde \lambda)\Phi_{,\tilde\lambda}}{r_{,\tilde\lambda}},
\end{eqnarray}
\end{subequations}
which yields 
\begin{subequations} 
\begin{eqnarray}
0
&=& 
 r_{,\tilde \lambda\tilde\lambda}
 +\frac{1}{2}\kappa r (\tilde\Phi_{,\tilde\lambda})^2,
 \\
Z_{,\tilde\lambda} 
&=&
0, \label{Zode}\\
{L}_{,\tilde\lambda}&=&
      \frac{ (\tilde\lambda + Z)\tilde\Phi_{,\tilde\lambda}}{r } ,
   \\
V_{,\tilde\lambda\tilde\lambda} &=&
 -\frac{1}{\tilde\lambda}\left(\frac{\tilde\lambda^ 2 }{r^2 }\right)_{,\tilde\lambda}
  +\frac{ 2 Z r_{,\tilde\lambda}}{r^3}
   + \frac{\kappa }{r }\tilde\Phi_{,\tilde\lambda} {L} \label{hyp}  \\
\tilde \Phi_{,\tilde u} &=& \frac{{L}}{ 2r } + \frac{1}{2}V\tilde\Phi_{,\tilde\lambda}\label{ev_eqn_Phi}.
\end{eqnarray}
\end{subequations}
The solution for \eqref{Zode} is trivially $Z=Z_0(u)$, this function of  integration  should be fixed by the boundary values, so we have
\begin{subequations} \label{eq:fe_an_physical_lambda}
\begin{eqnarray}
0
&=& 
 r_{,\tilde\lambda\tilde\lambda}
 +\frac{1}{2}\kappa r (\tilde \Phi_{,\tilde\lambda})^2,
 \\
{L}_{,\tilde\lambda}&=&
      \frac{ (\tilde\lambda + Z_0)\tilde\Phi_{,\tilde\lambda}}{r } ,
   \\
V_{,\tilde\lambda\tilde\lambda} &=&
 -\frac{1}{\tilde\lambda}\left(\frac{\tilde\lambda^ 2 }{r^2 }\right)_{,\tilde\lambda}
  +\frac{ 2 Z_0 r_{,\tilde\lambda}}{r^3}
   + \frac{\kappa }{r }\tilde\Phi_{,\tilde\lambda} {L},   \\
\tilde \Phi_{,\tilde u} &=& \frac{{L}}{ 2r } + \frac{1}{2}V\tilde\Phi_{,\tilde\lambda}.\label{ev_eqn_Phi}
\end{eqnarray}
\end{subequations}
For $\hat x \in [0,1]$, we introduce the grid function
\begin{equation}
\tilde \lambda = \frac{\hat x}{1-\hat x}\;\;,\;\;
\frac{d\tilde \lambda}{d \hat x}=\frac{1}{(1-\hat x)^2},
\end{equation}
where $\tilde \lambda(0)=$ and $\lim_{\hat x\rightarrow 1}\tilde \lambda = \infty$.
This means the physical domain of $\tilde \lambda$ is completely mapped onto the domain of $ \hat x$ and is  referred to as coordinate domain of  $\tilde \lambda$ is compactified. 

In terms of this new coordinate $\hat x$,  \eqref{eq:fe_an_physical_lambda} are
\begin{subequations} \label{eq:hier_coord_comp}
\begin{eqnarray}
0&=&r_{,\tilde u \tilde u} 
- \frac{V}{4r}\,(1 - \hat x)^2  \left[ r^2 (1 - \hat x)^2 V_{,\hat x} \right] _{,\hat x}
\nonumber\\
  &&
  + \frac{1}{2} (1 - \hat x)^2 \left[ r_{,\tilde u} V_{,\hat x} -  r_{,\hat x} V_{,\tilde u} \right] 
\nonumber\\
  &&
  + \frac{1}{2} \kappa\, r \left(\tilde  \Phi_{,\tilde u} \right)^2,\\
0
&=& 
 r_{,\hat x\hat x}
  -\frac{2r_{,\hat x}}{1-\hat x}
 +\frac{1}{2}\kappa r (\tilde \Phi_{,\hat x})^2,
 \\
{L}_{,\hat x}&=&
      \frac{ [\hat x + (1-\hat x)Z_0]\tilde \Phi_{,\hat x}}{(1-\hat x)r } ,
   \\
\Big[(1-\hat x)^2V_{,\hat x}\Big]_{,\hat x} &=&
 -\frac{1-\hat x}{\hat x}\left\{\frac{\hat x^ 2 }{[(1-\hat x)r]^2 }\right\}_{,\hat x}
  +\frac{ 2 Z_0 r_{,\hat x}}{r^3}
  \nonumber\\
  &&
   + \frac{\kappa }{r }\tilde \Phi_{,\hat x} {L}   ,\\
\frac{\tilde \Phi_{,\tilde u}}{1-\hat x} &=& \frac{{L}}{ 2r(1-\hat x) } + \frac{1}{2}V(1-\hat x)\tilde\Phi_{,x}.\label{ev_eqn_Phi}\label{ev_eqn_Phi}
\end{eqnarray}
\end{subequations}
This system of partial differential equations \eqref{eq:hier_coord_comp} has singularities at $\hat x=1$. These singularities can be absorbed into the regularized fields 
\begin{eqnarray}
R&=&(1-\hat x)r,\label{eq:r_reg_comp}\\
W&=&(1-\hat x)V,\label{eq:V_reg_comp}\\
\Phi&=&\frac{\tilde \Phi}{(1-\hat x)},\label{eq:phi_reg_comp}
\end{eqnarray}
which yields
\begin{subequations} \label{eq:hierachy_coord_comp_final}
\begin{eqnarray}
0
&=& 
 R_{,\hat x\hat x}
 +\frac{1}{2}\kappa R \Big[(1-\hat x) \Phi_{,\hat x}-\Phi\Big]^2,
 \\
{L}_{,\hat x}&=&
      \frac{ [\hat x + (1-\hat x)Z_0]\Big[(1-\hat x) \Phi_{,\hat x}-\Phi\Big]}{R } ,
   \\
W_{,\hat x\hat x}&=&
 -\frac{1}{\hat x}\left(\frac{\hat x^ 2 }{R^2 }\right)_{,\hat x}
  +\frac{ 2 Z_0}{R^3}  [(1-\hat x)R_{,\hat x}+R]
  \nonumber\\
  &&
   + \frac{\kappa }{R }  {L} \Big[(1-\hat x) \Phi_{,\hat x}-\Phi\Big],  \\
 \Phi_{,\tilde u} &=& \frac{1}{ 2 }\left\{ \frac{L}{R}+  W\Big[(1-\hat x) \Phi_{,\hat x}-\Phi\Big]\right\}.\label{eq:rhs_phys_reg_dphidu}
\end{eqnarray}
\end{subequations}
It is not difficult to see 
that the equations \eqref{eq:hierachy_coord_comp_final} are equivalent to those derived in the conformally compactified equations \eqref{eq:hierachy_conf_final}
%%%%%%%%%%%%%%%%%%%%%%%%%%%%%%%%%%%%%%%%%%%%%%
\section{Direct extraction of scalar News  function at null infinity}\label{sec:num}
In the study of gravitational collapse and scalar field dynamics near criticality, an important quantity is the scalar news function, which governs the radiative loss of mass through null infinity. Accurate extraction of this quantity is essential for tracking the decay of the Bondi mass and understanding the global properties of the spacetime. 
As an application of our previous findings about the partial differential equations in physical and conformal spacetimes, we revisit a key technical aspect of previous work \cite{Madler:2024kks}.

Therein, the news calculation on the compactified grid requires evaluating a second-order mixed $u\hat{x}$-derivative of the scalar field.
This numerical derivative was  noisy in \cite{Madler:2024kks}, due to the employed  spectral method and the e-folding \cite{Gundlach:2007gc} of the scalar field near the critical solution.
For this reason, the news function driving the Bondi mass decay was not extracted. 
However, the use of the regularized scalar field $ \Phi$ (see \eqref{eq:phi_reg_comp}) avoids the calculation of this mixed derivative. 
In fact the news function can be read off directly from the hypersurface fields given by the right hand side of \eqref{eq:rhs_phys_reg_dphidu} at the outer boundary.% at $\hat x=1$.

Taking these considerations into account, we now demonstrate how the numerical code of \cite{Madler:2024kks} can be adapted to compute the news function. 
The code solves the field equations with the origin, $\tilde \lambda = 0$, on to the central worldline of spherical symmetry, which implies that the metric boundary conditions satisfy the regularity requirements $r = \tilde\lambda + O(\tilde \lambda^3)$ and $V = 1 + O(\tilde \lambda^2)$. As a consequence, the second-order differential equation of the $V$-hypersurface equation can be cast into a first-order one for the new auxiliary field $Q = (V-1)/\tilde \lambda$.
A further consequence is $Z_0=0$. 
We evolve the  initial data  ${\tilde\Phi = \sqrt{12}\left[\arcsin(\frac{b\tilde\lambda}{a+\tilde\lambda})-\arcsin b\right]}$ with $a=2$ and $b\simeq 0.9472018$. The given parameter choice  corresponds to supercritical initial data evolving into a black hole spacetime for code time\footnote{For the employed initial data, the critical solution is found for  $b=b_c\simeq0.947201675$. The parameter $a$ is a scaling parameter. It determines the time, $u_c$, of formation of the critical solution as time $u_c=3a$. Black holes thus form for values $b>b_c$ and at times bigger than $u_c$, see \cite{Madler:2024kks}.} $\tilde u\gtrsim 6$.

\begin{figure}[t]
    \centering
    \includegraphics[width=0.45\textwidth]{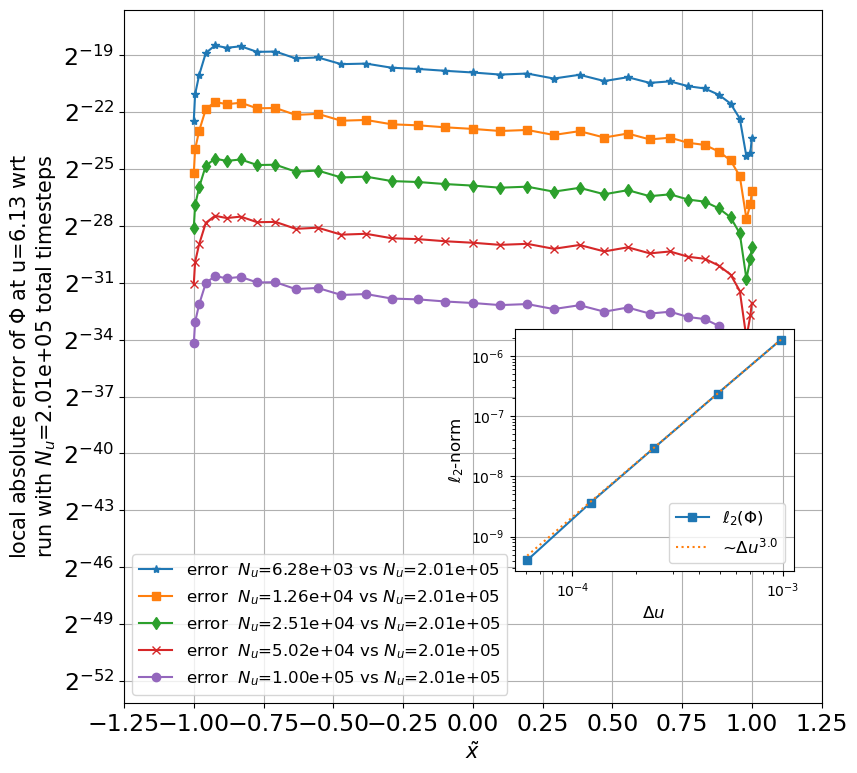}
    \caption{Absolute error of the scalar field $\Phi$ as a function of $\tilde x$ resulting from six different temporal resolution runs with $N_u\in\{{6.3\cdot10^3},{1.3\cdot10^4}, {2.5\cdot10^4},{5\cdot10^4}, 1\cdot10^5,2\cdot10^5\}$ time steps. All errors are calculated at the same time $u=6.13$. The curve with the blue stars, orange squares, green diamonds, red crosses and violet dots correspond  absolute error difference with respect to  the run having $2\cdot10^5$ time steps with  the run with $6.3\cdot10^3$,   ${1.3\cdot10^4}$,  ${2.5\cdot10^4}$, ${5\cdot10^4}$ and $1\cdot10^5$ time steps, respectively. The spatial number of grid points is kept fixed at $33$ collocation points.
    By doubling the number of temporal spacings the error decays by a factor of $8=2^ 3$.
    The inlet curve with blue squares shows the dependence of the $\ell_2$-norm at final time  as function of the time step employed in the simulations. 
    The orange dotted curve displays the expected behavior of the temporal discretization error $\sim\Delta u^ 3$. 
    The alignment of the $\ell_2$-norm values to this curve  confirms the third order convergence of the numerical scheme. }   
    \label{fig:convergence}
\end{figure}

The physical affine parameter is substituted by the grid function\footnote{This is the compactified grid function of \cite{Madler:2024kks} with the parameter choice $A=1$.} $\tilde \lambda = 2(1+\tilde x)(1-\tilde x)^{-1}$ with $\tilde x\in[-1,1]$ so that $\tilde\lambda(-1)=0$ and $\lim_{\tilde x\rightarrow1}\tilde\lambda(\tilde x)=\infty.$
The code of \cite{Madler:2024kks} uses the physical scalar field $\tilde \Phi$ and already employs the renormalized area distance $R = (1-\tilde x)r$.   
We change the implementation of \cite{Madler:2024kks} by using ${\Phi = (1-\tilde x)\tilde \Phi}$ instead of $\tilde \Phi$.
The physical relevant fields, $f\in (\Phi, R, L, Q, V)$, are discretized like $f( u_n, \tilde x_i) = \sum_{k=0}^N f_k(  u_n)T_k(\tilde x_i)$, where 
$N+1$ is the number of grid points discretizing equidistantly the half unit circle in $N$ arc segments, $ u_n = n \Delta  u$ with $\Delta  u=(N+1)^{-2}$ is the discretized representation of the proper time along the central geodesic, 
$\tilde x_i$ are Gauss-Lobatto points, ${\tilde x_i=-\cos(\frac{i\pi}{N})}$ with $i=0,...N$, which are the extrema of the Chebycheff polynomials $T_k(\tilde x)$ of first kind. 
Under these definitions the modified system %of \cite{Madler:2024kks} 
gives the set of hypersurface equations
\begin{subequations} \label{eq:hierachy_num}
\begin{eqnarray}
\tilde\Phi_{,\tilde x}&=&(1-\tilde x) \Phi_{,\tilde  x}-\Phi\\
0
&=& 
 R_{,\tilde x\tilde x}
 +\frac{1}{2}\kappa R \Big(\tilde \Phi_{,\tilde x}\Big)^2,
 \\
{L}_{,\tilde x}&=&
      \frac{ 2(1+\tilde x)\tilde  \Phi_{,\hat x}  }{R } ,
   \\
Q_{, \tilde x}&=& 
\frac{1}{(1+\tilde x)^2}
-\frac{4}{R^2}
+\frac{1}{2}\left(\frac{L}{1+\tilde x}\right)^2,\label{eq:dQdx_num} %\\
 %\label{eq:rhs_phys_reg_dphidu}
\end{eqnarray}
\end{subequations}
and the evolution equation
\begin{equation}\label{eq:dphidu_num}
    \Phi_{, u} = \frac{1}{ 2 }\left\{ \frac{L}{R}+  \frac{1}{4}[(1-\tilde x) +2(1+\tilde x)Q]\tilde \Phi_{,x}\right\}.
\end{equation}
The system is solved by presenting initial values for $\Phi$ on an initial null cone $ u=0$, while using the boundary conditions $R(u,-1)=0$, $R_{,\tilde x}(u, -1)=2$ and  $L(u,-1)=Q(u,-1)=0$, while $\Phi_{, u}(u,-1)=\Phi_{,\tilde x}(u,-1)$. 
With the initial data on the cone $ u=0$, the hypersurface equations \eqref{eq:hierachy_num} are solved sequentially. 
Then the right hand side of \eqref{eq:dphidu_num} can be solved on $u=0$. 
Given the time step $\Delta u = (N+1)^{-2}$, the scalar field is then calculated with the Shu and Osher third order stability \cite{1988JCoPh..77..439S} preserving Runge-Kutta scheme to find $\Phi(\Delta  u)$. 
The scalar field $\Phi(\Delta   u)$ on the cone $\Delta  u$ then serves as new initial data to obtain $\Phi(2\Delta  u)$.
This hierarchical hypersurface-evolution algorithm is repeated until the end of the simulation time.  
In Fig.~\ref{fig:convergence}, we show the convergence of spatial distribution of  the  absolute error (with respect to a high resolution run) of the scalar field. If $\Delta u$ is divided by a factor of two\footnote{This corresponds to  the number of time steps minus one being  doubled.} while keeping the final simulation time fixed, the error decays by a factor of $8=2^3$. This is consistent with third order convergence. 
The inlet in Fig.~\ref{fig:convergence} also confirms that the error of the time integration is $O(\Delta  u^3)$. 

For the massless Einstein-scalar field equations in spherical symmetry, the variables of physical interest at null infinity are: i) the redshift $H$, ii) the Bondi time $u_b$, iii) the scalar monopole $C$, iv) the Bondi mass $m_b$, and v) the news function $N$.
The function $H$ is numerically extracted at the outer boundary according to $H( u) = R( u, -1)/4$. Given $H$, the Bondi time is found by integration of \eqref{eq:conf_redshift}. Moreover, once $H$ is determined, the Bondi mass is found from $m_B( u) =H( u)- R_{,x}( u, 1)/2$.
Given $H$, the scalar monopole $C$ is found numerically by evaluation of the scalar field at $\tilde x=1$, i.e. $C = \Phi( u, -1)/(4H)$.
The news function $N$ is defined with respect to the Bondi time $\tilde u_b$ in \eqref{eq:def_news}. 
In terms of the code time $ u$, it is $N = H C_{,u} = HH_{, u}\Phi_{[0]} + H^2 \Phi_{[0], u}$ which follows from \eqref{eq:monopole}, \eqref{eq:conf_redshift} and \eqref{eq:def_news}. 
The derivative $\Phi_{[0], u}$ can be read from the right hand side of \eqref{eq:dphidu_num} evaluated at the outer boundary, while the derivative $H_{, u}$ is read from the right hand side of \eqref{eq:dQdx_num}. 
The latter is a consequence of $\lim _{\lambda\rightarrow \infty}Q = (\ln H^2)_{, u}$, see \cite{Crespo:2019mcv}.
With knowledge of the news function, Bondi mass and Bondi time, the asymptotic balance law \eqref{eq:BmassLoss_News} can be calculated to verify the consistency of the extracted quantities.

\begin{figure}[h]
    \centering
    \includegraphics[width=0.45\textwidth]{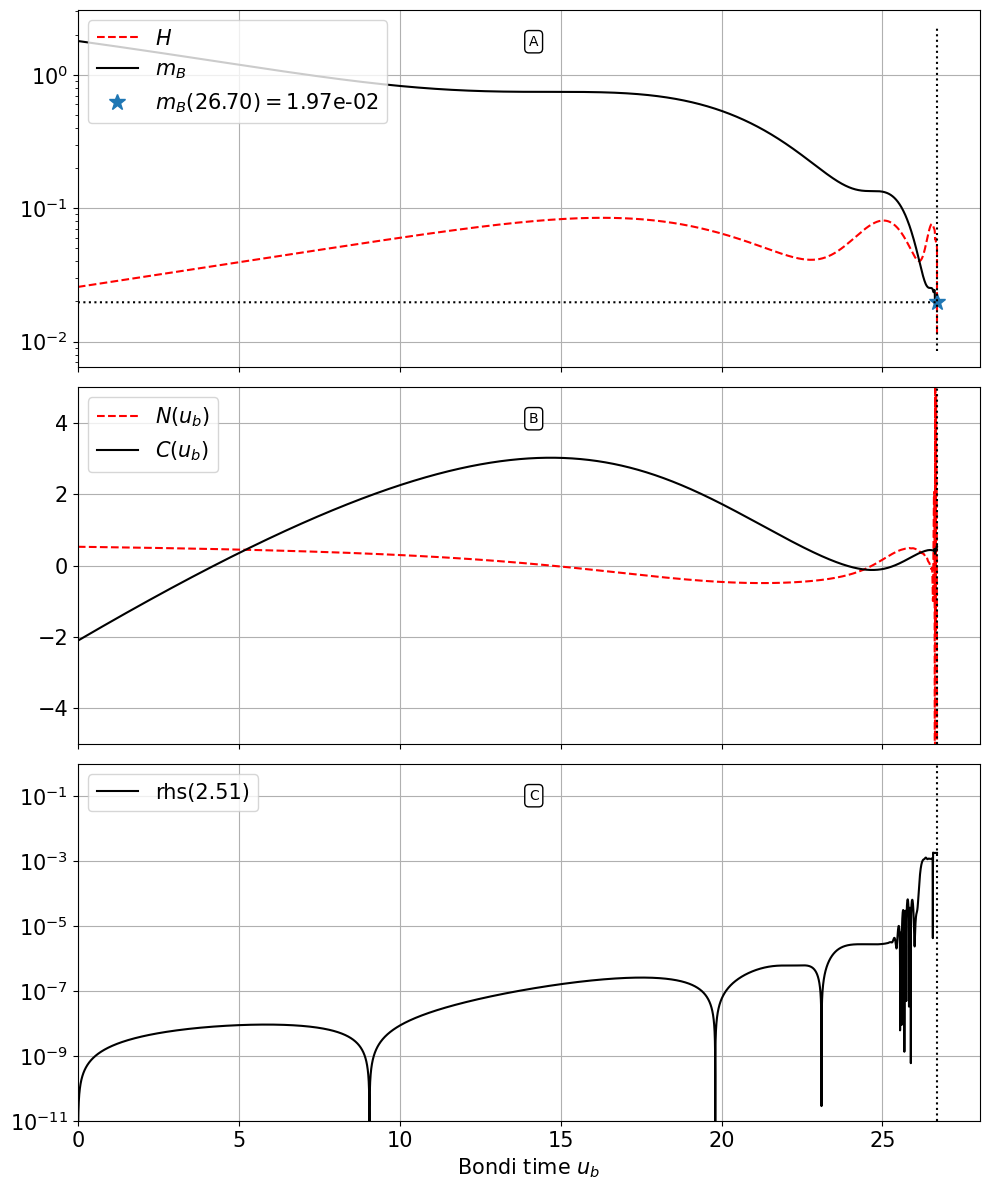}
    \caption{
    In all three panels, quantities are evaluated at null infinity, $\tilde x=1$, and the abscissa is the Bondi time $u_b$. The upper panel A shows the evolution of the redshift factor $H$ (red dashed line) and the Bondi mass $m_B$ (solid black line). The middle panel B shows the news function $N$ (dashed red line)  and the scalar monopole $C$ (solid black line). The lower panel C shows the  numerical  error of the calculation of the right hand side of \eqref{eq:BmassLoss_News}.  The vertical dotted lines correspond to the Bondi time of the last null cone connecting  with the central geodesic. The horizontal dotted line in panel A and the blue star point out the final value of the Bondi mass at the threshold of black hole formation.  }   
    \label{fig:plot}
\end{figure}

In Fig.~\ref{fig:plot}, we display the behavior of the asymptotically relevant variables, redshift $H$, Bondi mass $m_B$, scalar monopole $C$, news function $N$ and the balance law \eqref{eq:BmassLoss_News} as a function of the Bondi time $u_b$.
To our knowledge,  this is the first extraction of a news function in the context of critical phenomena while using affine-null coordinates. Up until now scalar field news functions have been extracted using Bondi-Sachs coordinates
by \cite{Purrer:2004nq, Barreto:2014rba,Barreto:2017kta}.
As explained in Sec.~\ref{sec:conf}, the formation of a black hole is indicated if the redshift $H\rightarrow0$. 
In panel A of Fig.~\ref{fig:plot}, it is seen that the redshift (red dashed line) drops to zero at the Bondi time $ u_H :=  u_b\sim26.70$.
This $\tilde u_b=const$ slice  corresponds to the last null cone that connects an asymptotic observer at null infinity with the central geodesic, where $\tilde x=-1$ and from where the system \eqref{eq:hierachy_num} and \eqref{eq:dphidu_num} 
is integrated towards $\tilde x=1$. 
Beyond the time $u_H$, an asymptotic observer has no information on how the scalar field continues to collapse toward the central geodesic.
\begin{figure}[h]
    \centering
    \includegraphics[width=0.45\textwidth]{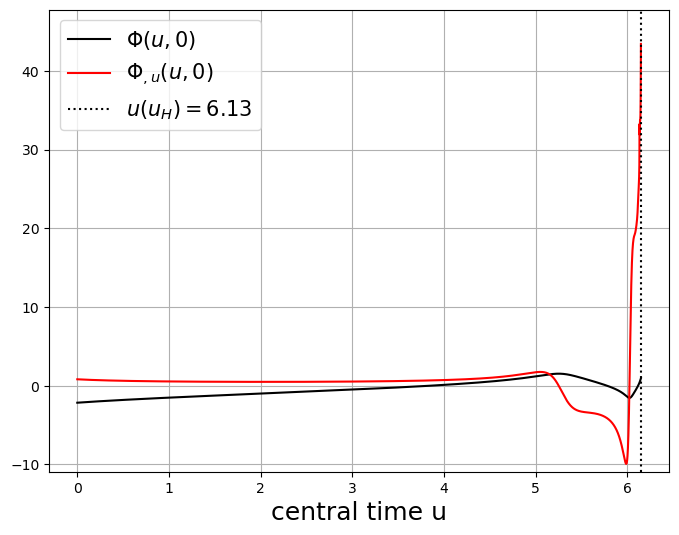}
    \caption{Evolution of the scalar field and its $u-$derivative at the origin in the compactified coordinate where $\tilde \lambda=0$. 
    The dotted line corresponds to the code time of the last null cone connecting to $\mathcal{I}$.}
    \label{fig:phi}
\end{figure}
Panel A also displays the (exponential) decay of the Bondi mass with a final value (blue star) of $m_B=1.97\cdot10^{-2}$  measured at $\mathcal{I}$.
The news function (red dashed line) and the scalar monopole (black solid line) are shown in panel B of Fig.~\ref{fig:plot}.
In the behavior of the two functions one can see that the news function is the derivative of the monopole with respect to the Bondi time $u_b$.
It can be understood by correlating the zeros of the news function with the extrema of the scalar monopole. 
We observe in panel B that they coincide.
The Bondi mass decay is coupled with the square of the news function via \eqref{eq:BmassLoss_News}. 
In panel B of Fig.~\ref{fig:plot}  the right hand side of \eqref{eq:BmassLoss_News} is shown  using the data of the Bondi mass displayed in panel A and the news function in panel B. 
This balance law was also shown graphically by displaying its components by Barreto \cite{Barreto:2014rba} in physical space.  
We can see that the balance law holds up to numerical precision.
 For about 80\% of the overall simulation time of $u_H$, the absolute error in the balance law is $\sim 10^{-7}$. 
It then increases to about $10^{-3}$ until the simulation crashes due to black hole formation. 
The increase of error can be understood by looking at the evolution of the scalar field at the origin, i.e along the central geodesic.
Thereat are the start points of the null rays  emanating to null infinity. 
The scalar field (black curve) and its $u$-derivative (red dashed curve) both evaluated at the origin are shown in Fig.~\ref{fig:phi}. 
The abscissa is the proper time at the central geodesic and it is also the time measured in the numerical simulation. 
The central time determines the Bondi time in Fig.~\ref{fig:plot} by integrating \eqref{eq:conf_redshift}. 
We observe that the zero crossings of the scalar field decrease exponentially towards the final time.
The central time $u=6.13$ has   the Bondi time $u_H$ at $\mathcal{I}$, which is the time when the last null cone emanating from the central is connected with null infinity. 
The exponential decrease in the zero crossings of the scalar field towards final time, also means that the $u$-derivative of the scalar field along the central geodesic exponentially in amplitude towards the final time (see the red curve in Fig.~\ref{fig:phi}).
As the $u$-derivative of the scalar field along the geodesic is determined by a $\tilde{\lambda}$-derivative evaluated at\footnote{The system \eqref{eq:fe_an_physical_lambda} and the regularity conditions for a freely falling observer at the central geodesic imply $\Phi_{,\tilde u}|_{\tilde \lambda=0} = \Phi_{,\tilde \lambda}|_{\tilde \lambda=0}
$. Consequently, in code variables $\Phi_{, u}|_{\tilde x=-1} =\Phi_{,\tilde x}|_{\tilde x=-1}$, where the spatial derivative is taken over the data on a given cone $u=const$.} $\tilde \lambda=0$ , for times close to black hole formation time,  the spatial derivatives at the vertex also become stiffer with increasing amplitudes.
As consequence high numerical resolution is required at the vertex to capture the scalar fields dynamical behavior along the geodesic.
It is this stiffness and because of  the employed spectral methods %in %the code of \cite{Madler:2024kks}, 
that leads to the increase of error towards $10^{-3}$ in panel C of Fig.~\ref{fig:plot}  and  spoils the numerical accuracy in the evaluation of the balance law at times before black hole formation.
Despite the loss of accuracy, the temporal convergence (see Fig.~\ref{fig:convergence}) on the final null cone $u_H$ indicates the numerical consistency/convergence of the results. 
%%%%%%%%%%%%%%%%%%%%%%%%%%%%%%%%%%%%%%%%%%%%%%

\section{Summary and discussion}\label{sec:discuss}
The Einstein-scalar field equations for a massless scalar field in spherical symmetry is a simple dynamical system to capture all the important aspects of radiative fields in General Relativity. 
In this work, we present its underlying partial differential equations in the physical spacetime and in a conformal unphysical spacetime, 
while charting  both spacetimes with  affine-null coordinate systems.
Apart from two coordinates  associated to spherical symmetry, in an affine-null coordinate system,  one coordinate is labelling a family of null hypersurfaces and another coordinate is an affine parameter that parameterizes the surface forming rays.
In contrast to the classical Bondi-Sachs coordinates, the system of partial differential equations  is not automatically given by a natural   hierarchical system that can be straighforwardly integrated with an initial data profile on a null hypersurface.
This is due to a time derivative of the area distance in one of the hypersurface equations. 
However, the hierarchy can be restored by introducing new fields.
This is possible for the conformal Einstein equations as well as in those of the physical manifold. 
In fact, our work is the first in presenting such a hierarchy for the conformal Einstein-scalar field equations in affine-null coordinates, whose derivation and involved variables are significantly less obvious compared to the same equations in physical space like in \cite{Crespo:2019mcv}.
It turns out that the conformal Einstein equations are naturally regular at null infinity and this regularity results from the choice of the conformal factor.
The Einstein equations in physical space, however, are singular for large values of the affine parameter.
To map this infinity onto a numerical grid, the radial coordinate is compactified using a typical coordinate compactification, $\tilde\lambda = \hat x/(1-\hat x)$, similar to the one used in \cite{Madler:2024kks}. 
The start point  $\hat x=0$   is on the central geodesic, while $\hat x=1$ corresponds to the limit $\tilde \lambda \rightarrow \infty$.
The resulting system of partial differential equations are also singular  at $\hat x=1$, and thus are unsuited for numerical work.
However, introducing the new regularized fields for the metric fields mends the system of equations from these coordinate singularities. 
Moreover,  we observe that the compactified system in physical space with properly  regularized field for the metric, i.e.\eqref{eq:r_reg_comp} and \eqref{eq:V_reg_comp},  and  scalar field, i.e. \eqref{eq:phi_reg_comp}, is {\it identical} with the hierarchy  derived from the conformal (unphysical) field equations.
This equality of the field equation in spherical symmetry  holds for affine-null coordinates  as well as for Bondi-Sachs coordinates, the latter is demonstrated for completeness in App.~\ref{sec:appendix}. 
Therefore to model the behavior of a radiative scalar field in spherical symmetry, it is sufficient to work in physical space  while introducing a compactified radial coordinate and numerically renormalize the resulting field equations so that they are free of  coordinate singularities `at the location of infinity'.

In \cite{Madler:2024kks}, a partially   regularized set of equations in physical space was used to show the echoing of the critical phenomena for the redshift and the corresponding decay of the Bondi mass.
The reason for using a partially regularized set of equation was that the compactified Einstein equations in physical space, corresponding to \eqref{eq:hier_coord_comp}, in principle only require a regularization for the metric fields.
This is because the compactified evolution equation, \eqref{ev_eqn_Phi}, can be easily regularized by moving the singular term at $\hat x=1$ from the  left hand side to the right hand side.
As a result, the time derivative of the scalar field at  'compacified infinity', $\hat x=1$, is zero.
The initial  scalar field also vanishes on $\hat{x}=1$, due to the Sommerfeld radiation condition \eqref{eq:SommerFeld}. 
The scalar field is therefore always zero on $\mathcal{I}$.
Consequently, the monopole can only be numerically determined by calculation of a derivative at $\mathcal{I}$.
The news function responsible for the Bondi mass decay was not extracted in \cite{Madler:2024kks}, in Sec.\ref{sec:num} we show how this issue can be resolved through a modification of its scheme. The key difference in the new approach is that the regularized scalar field \eqref{eq:phi_reg_comp} is incorporated into the existing code, thereby enabling a more accurate extraction of the news function; see Fig.\ref{fig:plot}.

Thanks to our successful derivation of hierarchical evolution hypersurface equations for the affine-null metric formulation of the Einstein-scalar field equations in the conformal space, we have been able to naturally find regularized fields by making use of affine-null coordinates derived both in an unphysical conformally compactified spacetime and in a coordinate-compactified physical spacetime. Due to the equivalence we have demonstrated with the hierarchical equations in the affine-null formulation of the physical spacetime for spherical symmetry and a scalar field, it would be highly interesting to verify if such an equivalence can also be established for systems with less symmetry than spherical symmetry and/or including other matter fields. This would help in naturally finding regularized fields with well-behaved properties at null infinity.
Starting points for such investigations, could be the articles \cite{Winicour:2012znc,Madler:2018bmu}, those works already showed the existence of hierarchies for the affine-null coordinates. 
It seems relatively straighforward (but may be technically tedious) to extend our presented procedure for this purpose.  Work into this direction is already in progress.

%%%%%%%%%%%%%%%%%%%%%%%%%%%%%%%%%%%%%%%%%%%%%%%%%%%%%%%%%%%%%%%%%%%%%%%%
\begin{acknowledgements}
E.G. acknowledges financial support from CONICET and SeCyT-UNC.
T.M. thanks  the organizers of the ``Theoretical High-Energy Physics Workshop on Gravity and Holography'' in Valparaiso/Vi\~na de Mar, Chile where some of this work was presented. E.G thanks Amato for enlightening discussions.
\end{acknowledgements}

\begin{appendix}
\section{Conformal and coordinate compactification with Bondi-Sachs coordinates}\label{sec:appendix}

In this section, we repeat for completness the analysis in Bondi-Sachs coordinates. 

The conformal Bondi-Sachs coordinates are $y^a = (u, y, y^a)$ and the  Bondi-Sachs coordinates in the physical spacetime are $\tilde y^a = (\tilde u, \tilde r, \tilde y^A)$. 
The physical space coordinates are defined at the central geodesic like the affine null coordinates $\tilde x^a$, respectively.
Indeed, $\tilde y^0 = \tilde u$ and $\tilde y^A$ carry over the same meaning.
However,  the radial coordinate $\tilde y^1=\tilde r$ in an area distance so that $\det(\tilde g_{AB}) = \tilde r^2\mathfrak{q}$.
The physical Bondi-Sachs metric takes its classical form in spherical symmetry in outgoing  null coordinates
\begin{equation}\label{eq:BS_metric_phys}
ds^2_{BS} = -\frac{V}{\tilde r}e^{2\beta}d\tilde u^2 -2e^{2\beta}d\tilde u d\tilde r + \tilde r^2 q_{AB}d\tilde y^A d\tilde y^a.
\end{equation}
%%%%%%%%%%%%%%%%%%%%%%%%%%%%%%%%%%%
\subsection{Conformal field equations in unphysical spacetime}

In this section, we consider the field equations \eqref{eq:FE_conf_an} for the metric \eqref{eq:BS_metric_phys}.

 The transition from the physical spacetime to the conformally unphysical spacetime is given by 
\begin{equation}
g_{ab}(y^c) = \Omega^2(y^c)\tilde g_{ab}(y^c)\;\;,\;\;
\Phi(y^c) = \frac{\tilde \Phi(y^c)}{\Omega(y^c)}\;\;.
\end{equation}
By restriction of the coordinate chart $y^a$ to $\tilde y^a$,  we may equally consider the following situation
\begin{equation}
g_{ab}(\tilde y^c) = \Omega^2(\tilde y^c)\tilde g_{ab}(\tilde y^c)\;\;,\;\;
\Phi(\tilde y^c) = \frac{\tilde \Phi(\tilde y^c)}{\Omega(\tilde y^c)}\;\;.
\end{equation}
In this situation, i.e. in physical coordinates $\tilde y^a$, we take the conformal factor to be 
\begin{equation}
\Omega(\tilde y^a) = \frac{1}{1+\tilde r}.
\end{equation}
It is finite for $\tilde r=0$ and vanishes for  $\tilde r\rightarrow \infty$.
Using the physical Bondi-Sachs metric \eqref{eq:BS_metric_phys},  a conformal Bondi-Sachs metric is given by 
\begin{widetext}
\begin{equation}
\begin{split}
ds^2 =& \Omega^2 d\tilde s^2 = \frac{d\tilde s^2}{(1+\tilde r)^2} 
= -\frac{V}{\tilde r (1+\tilde r)^2}e^{2\beta}d\tilde u ^2 -2\frac{e^{2\beta}}{(1+\tilde r)^2}d\tilde u d\tilde r +  \left(\frac{\tilde r}{1+\tilde r}\right)^2q_{AB}d\tilde y^A d\tilde y^B\\
=& -\frac{V}{\tilde r(1+\tilde r)^2}e^{2\beta}d\tilde u ^2 -2 e^{2\beta} d\tilde u d\left(\frac{\tilde r}{1+\tilde r}\right)  +  \left(\frac{\tilde r}{1+\tilde r}\right)^2q_{AB}d\tilde y^A d\tilde y^B.
\end{split}
\end{equation}
\end{widetext}

We define the conformal Bondi-Sachs coordinates via the physical coordinates like 
\begin{equation}
y^a= (u,y,y^A) = \left(\tilde u, \frac{\tilde r}{1+\tilde r}, \tilde y^A\right).
\end{equation}
This  conformal coordinate definition gives
\begin{equation}
\tilde r(y) = \frac{y}{1-y},
\end{equation}
so that $r(0) = 0$ and $\lim_{y\rightarrow 1}r(y) = \infty$, meaning that the boundary $\mathcal{I}$ is at $y=1$.
Indeed the conformal factor $\Omega$ in  conformal Bondi-Sachs coordinates becomes
\begin{equation}\label{eq:Omega_BS_cong}
\Omega(y) =1-y\;\Rightarrow\;\Omega|_{y=1}=0\;,\;(\nabla_a \Omega)dy^a|_{y=1} = -dy\neq0\;,
\end{equation}
%}
while the conformal metric reads
\begin{equation}
ds^2_{BS} 
= -y^{-1} (1-y)^3 Ve^{2\beta}d u ^2 -2 e^{2\beta} d u dy + y^2 q_{AB}d y^A d y^B.
\end{equation}
Its volume element is $\sqrt{-g} =y^2 e^{2\beta}\sqrt 
\mathfrak{q}$ and the inverse metric components
\begin{equation}\label{eq:gab_up_BS_conf}
\begin{split}
&g^{ u  x } = -e^{-2\beta}\;\;,\;\;
g^{yy } = y^{-1} (1-y)^3 V e^{-2\beta}\\
&
g^{ A  B } = y^{-2} q^{ A  B }.   
\end{split}
\end{equation}
The conformal metric has   the property 
\begin{equation}
ds^2_{BS} \Big|_{y=1}
=  0\cdot du^2  + q_{AB}d y^A d y^B
\end{equation}
showing that the conformal metric at $y=1$ is a three-dimensional  null metric since $g_{yy}=0$. 
The relevant Ricci tensor components to form the field equations with respect to  the conformal metric are
\begin{widetext}    
\begin{align}
R_{uu}=& y^3 V\left[2\beta_{,uy}+ \frac{3}{2}(y^2V)_{,y} + V(y^3\beta_{,y})_{,y}
+\frac{1}{2}e^{-2\beta}(y^3V_{,y}e^{2\beta})_{,y}\right]\\
R_{yy}=&\frac{4\beta_{,y}}{y}\\
R_{AB}=&\left\{1-[(1-y)^3V ]_{,y}e^{-2\beta}\right\}q_{AB}
\end{align}
For the field equations \eqref{eq:FE_conf_an} we have
\begin{align}
0=&R_{ab} 
+\frac{2\Gamma^y_{ab}}{1-y} +g_{ab}\left[-\frac{(1-y)^2(x V)_{,y}e^{-2\beta}}{y^2}\right]
-\kappa  \left[(1-y) \Phi\right]_{,a}\left[(1-y) \Phi\right]_{,b},\\
0=&(1-y)^3 \nabla_a\nabla^a\Phi
+ (1-y)^2\Phi \nabla_a \nabla^a\Omega
-2 (1-y) \Phi  g^{ yy}   .
\end{align}
\end{widetext}
The relevant equations to form an evolution algorithm are
\begin{subequations}\label{eq:pde_conf_BS}
\begin{eqnarray}
\beta_{,y}&=&\frac{\kappa}{4}y(1-y)  \left[(1-y) \Phi_{,y}-\Phi\right]
^2,\\
V_{,y}&=&
\frac{e^{2\beta}}{(1-y)^2},\label{eq:V_eqn_BS_conformal}\\
( x\Phi)_{, u y} 
&=&
\frac{1- y }{2  y} \left\{ y(1- y)V [ (1-y)  \Phi_{, y}-  \Phi]\right\}_{, y}.\nonumber\\
\end{eqnarray}
\end{subequations}
We observe that the $V_{,y}$ equation in \eqref{eq:pde_conf_BS}  is not defined at the conformal boundary, $y=1$. 
We  postpone its regularization for the next section, where the physical space field equations are considered. 
%%%%%%%%%%%%%%%%%%%%%%%%%%%%%%%%%%%%%
\subsection{Regularized field equations in physical spacetime with coordinate compactification}
In physical spacetime the relevant components of the field equations for \eqref{eq:BS_metric_phys} are
\begin{align}
0=&
-\frac{2V}{\tilde r^2}(\tilde r\beta_{,\tilde u})_{,\tilde r}
+\frac{V }{2\tilde r^3}\left[e^{-2\beta}(\tilde rVe^{2\beta})_{,\tilde r}\right]_{,\tilde r}
+\frac{V_{,\tilde u}}{\tilde r^2}
\nonumber\\
&-\frac{VV_{,\tilde r}}{\tilde r^3}
-\kappa \left(\tilde \Phi_{,\tilde u}\right)^2,\\
\beta_{,\tilde r}=& \frac{\kappa \tilde r}{4}\left(\tilde \Phi_{,\tilde r}\right)^2,\label{eq:phys_EE_rr}\\
V_{,\tilde r}=&e^{2\beta},\label{eq:phys_EE_trAB}
\\
2\tilde r(\tilde r\tilde \Phi)_{,\tilde u\tilde r} = &  (\tilde rV\tilde \Phi_{,\tilde r})_{,\tilde r}.\label{eq:phys_wave}
\end{align}
The Misner-Sharp mass is given by (see eq. \eqref{eq:MisnerSharp_phys}) \begin{equation}\label{eq:mass-ms-bondi-sachs}
\tilde{M}=\frac{\tilde r}{2}\left(1-\frac{V}{\tilde r}e^{-2\beta}\right).
\end{equation}
A characteristic evolution algorithm consists of the three equations \eqref{eq:phys_EE_rr}, \eqref{eq:phys_EE_trAB} and \eqref{eq:phys_wave}.
For $ \hat y\in[0,1]$, consider the  coordinate transformation
\begin{equation}
\tilde r(\hat y) = \frac{ \hat y}{1- \hat y},
\end{equation}
which gives $\tilde r(0)=0$ and $\lim_{ \hat y\rightarrow 1}\tilde r( \hat y) = \infty$. 
Its derivative $\tilde r_{, \hat y} =  (1- \hat y)^{-2}$ is everywhere finite for $ \hat y\in[0,1)$ and  the first $\tilde r-$derivative of the a field $f(\tilde r)$ transforms like
$ f_{,\tilde r} = (1-\hat y)^2 f_{, \hat y}$,
whose application to the set of relevant equations, \eqref{eq:phys_EE_rr}, \eqref{eq:phys_EE_trAB} and \eqref{eq:phys_wave}, gives
\begin{subequations}
\begin{eqnarray}
\beta_{,\hat y}&=& \frac{\kappa \hat y(1- \hat y)}{4}\left(\tilde \Phi_{, \hat y}\right)^2,\\
V_{, \hat y}&=&\frac{e^{2\beta}}{(1- \hat y)^2},\label{eq:V_BS_phys}\\
\left(\frac{ \hat y\tilde \Phi}{1- \hat y}\right)_{,\tilde u  \hat y} 
&=&
\frac{1- \hat y }{2  \hat y} \left[ \hat y(1- \hat y)V\tilde \Phi_{, \hat y}\right]_{, \hat y},\label{eq:transf_phys_compact_ev_eqn}
\end{eqnarray}
\end{subequations}
in which the hypersurface equation \eqref{eq:V_BS_phys} and the evolution equation \eqref{eq:transf_phys_compact_ev_eqn} are both singular if $ \hat y=1$. 

The regularization of  $\tilde \Phi $ like $\tilde \Phi = (1-\hat y)\Phi$, where $ \Phi$ has a regular Taylor series at $\hat y = 1$ yields
\begin{subequations}\label{eq:reg_phi_corrd_phys}
\begin{eqnarray}
\beta_{,\hat y}&=& \frac{\kappa \hat y(1- \hat y)}{4}\left\{(1-\hat y) \Phi]_{, \hat y}\right\}^2,\\
V_{, \hat y}&=&\frac{e^{2\beta}}{(1- \hat y)^2},\label{eq:V_BS_phys_regPhi}\\
\left( \hat y \Phi \right)_{,\tilde u  \hat y} \label{eq:V_BS_phys_compact}
&=&
\frac{1- \hat y }{2  \hat y} \left\{ \hat y(1- \hat y)V[(1-\hat y) \Phi]_{, \hat y}\right\}_{, \hat y}.\label{eq:transf_phys_compact_ev_eqn_regPhi}
\end{eqnarray}
\end{subequations}

Comparison of \eqref{eq:reg_phi_corrd_phys} with \eqref{eq:pde_conf_BS} show that the two sets are identical, when setting $\hat y = y$. 
Thus the conformal space field equations and the field equations in physical with a compatified radial coordinate and renormalized fields are identical.
This  is the Bondi-Sachs equivalence between conformal space field equations and physical space field equations,  similar to  what was  shown in Sec.~\ref{sec:conf} and Sec.~\ref{sec:compact} for  affine null coordinates. 

In both systems of partial differential equations,    \eqref{eq:pde_conf_BS} and \eqref{eq:reg_phi_corrd_phys}, 
the $V$ hypersurface equations are still singular, either $y=1$ or $\hat y=1$. 
The equation \eqref{eq:V_BS_phys_regPhi} can be regularized with the Misner-Sharp mass  $\tilde M$ in physical space using  \eqref{eq:MisnerSharp_phys} which leads to 
\begin{equation}\label{eq:MS_hyp}
    \begin{split}
       0=&
       \tilde M_{, \hat y} 
        -\frac{\kappa \hat y(1-\hat y)}{2} \Bigl[(1- \hat y) \Phi_{, \hat y}- \Phi\Bigr]^2\tilde M 
\\
&+\frac{\kappa  \hat y^2}{4}\Bigl[(1- \hat y) \Phi_{, \hat y}- \Phi\Bigr]^2\;\;,
    \end{split}
\end{equation}
and its counterpart \eqref{eq:V_eqn_BS_conformal} in  conformal space is regularised using
\eqref{eq:Misner_Sharp_conf}, \eqref{eq:gab_up_BS_conf}, $\mathcal{R}=y$ and   and the conformal factor \eqref{eq:Omega_BS_cong}. 
The result of the latter will give an identical equation. 
Indeed, this regularized equation has already been used in numerical simulations by \cite{Purrer:2004nq} to study the critical phenomena at null infinity. In terms of $\tilde{M}$ the equation for the scalar field \eqref{eq:transf_phys_compact_ev_eqn_regPhi} reads
\begin{equation}
( \hat y\Phi)_{,\tilde u \hat y} 
=
-\frac{(1- \hat y) }{2 \hat y } \Bigl\{ \hat y [\hat y-2\tilde{M}(1-\hat y)]( \hat y  \Phi_{, \hat y}+  \Phi)e^{2\beta}\Bigr\}_{, \hat y}.
\end{equation}

%%%%%%%%%%%%%%%%%%%%%%%%%%%%%%%%%%%%%%
\section{Conformal affine-null version of the Schwarzschild metric}\label{App:sch}

The Schwarzschild metric expressed in Bondi (inertial) coordinates reads
\begin{equation}
d\tilde{s}^2=-\left(1-\frac{2m_b}{r_b}\right)d{u}_b^2-2d{u}_bdr_b-r_b^2q_{AB}d\tilde{x}^Ad\tilde{x}^B.
\end{equation}
Note that for the Schwarzschild metric $r_b$ is an affine parameter. Under the following coordinate transformation
\begin{eqnarray}
r_b&=&H\tilde\lambda+R_{[1]}+H,\\
du_b&=&\frac{d\tilde{u}}{H},
\end{eqnarray}
with $H$ and $R_{[1]}$ functions of $\tilde{u}$ and 
leaving the angular coordinates unchanged, we can express the Schwarzschild metric in general affine-null coordinates $(\tilde{u},\tilde \lambda,x^A)$
\begin{equation}
\begin{split}
d\tilde{s}^2=&-Wd\tilde u^2-2d\tilde{u}d\tilde\lambda-(H\tilde \lambda+R_{[1]}+H)^2q_{AB}d\tilde{x}^Ad\tilde{x}^B.
\end{split}
\end{equation}
with \begin{equation}
\begin{split}
    W(\tilde{u},\tilde\lambda)=&\frac{1}{H^2}
    -\frac{2m_b}{H^2(H\tilde\lambda+R_{[1]}+H)}\\&+\frac{2}{H}\left[(\tilde\lambda+1)\frac{dH}{d\tilde{u}}+\frac{dR_{[1]}}{d\tilde{u}}\right].
    \end{split}
\end{equation}
After a coordinate transformation using the compactified coordinate $x$ given by \eqref{eq:tildelambdax} (and $u=\tilde{u}$) we obtain the conformal metric $ds^2=\Omega^2d\tilde{s}^2$ as expressed in \eqref{eq:metric_an_conf} with
\begin{equation}
\mathcal{W}=\frac{\Omega}{H^2}\left(1-\frac{2m_b\Omega}{\mathcal{R}}\right)+\frac{2}{H}\frac{d\mathcal{R}}{du},
\end{equation}
where  $\Omega=1-x$ is the conformal factor and $\mathcal{R}=H+R_{[1]}\Omega$. If $m_b=0$, this metric reduces to the conformal Minkowski metric.
With these expressions at hand, it is easy to check that near the conformal boundary, the expansion of the conformal metric (in terms of powers of $\Omega$ up to the considered order) is given by \eqref{eq:schw_gen_conf} with $\mathcal{W}_0$ and $\mathcal{W}_1$ given by \eqref{eq:w0} and \eqref{eq:w1} respectively.

For completeness, we provide the expression for the conformal affine-null version of the Schwarzschild metric in Israel coordinates \cite{Israel:1966zz}. The physical metric is given by
\begin{equation}
\begin{split}
d\tilde{s}^2=&\frac{-2\tilde\lambda^2}{\tilde{u}\tilde{\lambda}-8m_b^2}d\tilde{u}^2-2d\tilde{u}d\tilde{\lambda}\\&
+\left(-\frac{\tilde{u}\tilde\lambda}{4m_b}+2m_b\right)^2q_{AB}d\tilde x^Ad\tilde x^B.
\end{split}
\end{equation}
The conformal version is determined by the fields
\begin{eqnarray}
    \mathcal{W}&=&\frac{2x^2}{(8m_b^2+u)x-8m_b^2},\\
   \mathcal{R}&=&H+R_{[1]}(1-x),
\end{eqnarray}
with
\begin{eqnarray}
     H&=&-\frac{u}{4m_b},\\
    R_{[1]}&=&\frac{u}{4m_b}+2m_b.
\end{eqnarray}
This is an alternative version to the conformal metric presented in \cite{Gallo:2021jxt} (in that reference, $\tilde{u}$ is denoted by $w$ and $\tilde\lambda$ by $y$).

\end{appendix}

\bibliographystyle{apsrev4-1}      
\bibliography{bib_compactified}

\end{document}